\documentclass[conference]{IEEEtran}
\IEEEoverridecommandlockouts
\usepackage{cite}
\usepackage{amsmath,amssymb,amsfonts}
\usepackage{algorithmic}
\usepackage{graphicx}
\usepackage{textcomp}
\def\BibTeX{{\rm B\kern-.05em{\sc i\kern-.025em b}\kern-.08em
    T\kern-.1667em\lower.7ex\hbox{E}\kern-.125emX}}

\newcommand{\extended}[1]{#1}
\newcommand{\paper}[1]{#1}
\renewcommand{\paper}[1]{}

\usepackage{colortbl}

\usepackage{varioref}

\usepackage{multirow, hhline}

\usepackage{adjustbox}
\usepackage{subfig}
\usepackage{graphicx}

\usepackage{bm}

\newcommand{\tamFunc}[1]{\mathtt{#1}} 
\newcommand{\sign}{\tamFunc{sign}} 
\newcommand{\vfyAgg}{\tamFunc{vfyAgg}}
\newcommand{\agg}{\tamFunc{agg}}

\newcommand{\pk}{\tamFunc{pk}}

\newcommand{\compFunc}[1]{\mathsf{#1}} 
\newcommand{\signComp}{\compFunc{sign}}
\newcommand{\vfyComp}{\compFunc{vfy}}
\newcommand{\genComp}{\compFunc{gen}}
\newcommand{\vfyAggComp}{\compFunc{vfyAgg}}
\newcommand{\aggComp}{\compFunc{agg}}

\newcommand{\encComp}{\ensuremath{\compFunc{enc}}}
\newcommand{\certComp}{\ensuremath{\compFunc{cert}}}

\newcommand{\signature}{\mathcal{S}}
\newcommand{\aggsignature}{\mathcal{SA}}

\newcommand{\oas}{\ensuremath{\compFunc{OAS}}}
\newcommand{\oaskgen}{\ensuremath{\compFunc{OAS.}\genComp}}
\newcommand{\oasaggpk}{\ensuremath{\compFunc{OAS.aggPK}}}
\newcommand{\oassgn}{\ensuremath{\compFunc{OAS.}\signComp}}
\newcommand{\oasaggsgn}{\ensuremath{\compFunc{OAS.}\aggComp}}
\newcommand{\oasvfy}{\ensuremath{\compFunc{OAS.}\vfyAggComp}}

\newcommand{\V}{\ensuremath{\mathcal{V}}}
\newcommand{\skV}{\ensuremath{sk_\V}}
\newcommand{\pkV}{\ensuremath{pk_\V}}

\newcommand{\Ow}{\ensuremath{\mathcal{O}}}
\newcommand{\skO}{\ensuremath{sk_\Ow}}
\newcommand{\pkO}{\ensuremath{pk_\Ow}}

\newcommand{\tokenrequest}{\emph{Token Request}}
\newcommand{\attestation}{\emph{Attestation}}

\newcommand{\actionFact}[1]{\mathrm{#1}}

\newcommand{\Honest}{\actionFact{Honest}}

\newcommand{\true}{\mathsf{true}}
\newcommand{\false}{\mathsf{false}}

\newcommand{\Z}{\mathbb{Z}}

\newcommand{\nicetilde}{\raisebox{0.5ex}{\texttildelow}}

\usepackage[dvipsnames]{xcolor}

\usepackage{listings}
\lstdefinestyle{mystyle}{
		literate=
*{~} {{{\nicetilde}}}{1}
{'} {{\textquotesingle}}{1},
	commentstyle=\color{brown},
	keywordstyle=\bfseries\color{blue},
	keywordstyle=[2]\bfseries\color{RedViolet},
	keywordstyle=[3]\bfseries\color{brown},
	keywordstyle=[4]\bfseries\color{RedViolet},
	numberstyle=\tiny\color{gray},
	basicstyle=\ttfamily\footnotesize,
	breakatwhitespace=false,         
	breaklines=true,                 
	captionpos=b,                    
	keepspaces=true,                 
	numbers=left,                    
	numbersep=5pt,                  
	showspaces=false,                
	showstringspaces=false,
	showtabs=false,                  
	tabsize=2
}
\lstdefinelanguage{Tamarin}{
	keywords={theory, begin, end, signing, multiset},
	otherkeywords={=, |, &, ], --, ->, -->, [, !, @, $, ==>},
	morekeywords=[2]{!, @},
	morekeywords=[2]{builtins, functions, equations},
	morekeywords=[3]{$},
	morekeywords=[4]{rule, restriction, lemma, Fr, In, Out, let, Ex, All, not},
	morecomment=[l]{//},
	morecomment=[s]{/*}{*/},
}

\usepackage{dashbox}
\usepackage{msc}

\usepackage[nolist]{acronym}

\usepackage{amsthm}

\newtheorem{tamarinrestriction}{Restriction}
\newtheorem{definition}{Definition}
\newtheorem{theorem}{Theorem}

\usepackage{hyperref}

\begin{document}

\title{One For All: Formally Verifying Protocols \\which use Aggregate Signatures \extended{(extended version)}}

\begin{acronym}
	\acro{BLS}[BLS]{Boneh-Lynn-Shacham}
	\acro{EUF-CMA}[EUF-CMA]{existential unforgeability under a chosen message attack}
	\acro{DEO}[DEO]{Destructive Exclusive Ownership}
	\acro{IETF}[IETF]{Internet Engineering Task Force}
	\acro{co-CDH}[co-CDH]{Computational Co-Diffie-Hellman}
	\acro{PoP}[PoP]{Proof of Possession}
	\acro{PKI}[PKI]{public key infrastructure}
\end{acronym}

\author{
\IEEEauthorblockN{Xenia Hofmeier$^{1}$, Andrea Raguso$^{1}$, Ralf Sasse$^{1}$, Dennis Jackson$^{2}$, and David Basin$^{1}$}
\IEEEauthorblockA{$^{1}$\textit{Department of Computer Science},
\textit{ETH Zurich},
Zurich, Switzerland \\
\{xenia.hofmeier, andrea.raguso, ralf.sasse, basin\}@inf.ethz.ch}
\IEEEauthorblockA{$^{2}$\textit{Mozilla}, \textit{UK} \\
research@dennis-jackson.uk}}

\maketitle

\begin{abstract}
Aggregate signatures are digital signatures that compress multiple signatures from different parties into a single signature, thereby reducing storage and bandwidth requirements. 
BLS aggregate signatures are a popular kind of aggregate signature, deployed 
by Ethereum, Dfinity, and Cloudflare amongst others, currently undergoing standardization
at the IETF. 
However, BLS aggregate signatures are difficult to use correctly, with nuanced requirements that must be carefully handled by protocol developers. 

In this work, we design the first models of aggregate signatures that enable formal verification tools, such as Tamarin and ProVerif, to be applied to protocols using these signatures. 
We introduce general models that are based on the cryptographic security definition of generic aggregate signatures, allowing the attacker to exploit protocols where the security requirements are not satisfied. 
We also introduce a second family of models formalizing BLS aggregate signatures in particular. 
We demonstrate our approach's practical relevance by modelling and analyzing in Tamarin a device attestation protocol called SANA. 
Despite SANA's claimed correctness proof, with Tamarin we 
uncover undocumented assumptions that, when omitted, lead to attacks.

\end{abstract}

\section{Introduction}

Digital signatures are well established and widely used. 
Aggregate signatures are an extension of traditional digital signatures which have found favor in applications where many independent parties interact, such as the Ethereum blockchain \cite{eth2bookUpgradingEthereum}, Cloudflare's distributed random beacon \cite{drandCryptography}, and Dfinity's `Internet Computer' \cite{internetcomputerChainkeySignatures}. 
Aggregate signatures were first introduced as a cryptographic primitive in \cite{bonehAggregateVerifiablyEncrypted2003a} as an extension of Boneh-Lynn-Shacham (BLS) signatures \cite{boneh2001short}.
BLS aggregate signatures are the most popular instantiation of aggregate signatures and they are undergoing standardization at the IETF~\cite{bonehDraftirtfcfrgblssignature05BLSSignatures2022}.

Unfortunately, naive constructions of BLS aggregate signatures are vulnerable to signature forgeries 
using so-called \emph{rogue public key attacks}, see Section~\ref{Sec:Background:BLS},
violating the basic security definition of aggregate signatures~\cite{bonehGraduateCourseApplied}.
Different mitigations have been proposed to address this issue, but unfortunately none of them is suitable in all circumstances. So the current IETF draft requires implementors to choose between mitigations. 
This leaves protocols open to attack if implementors select the wrong mitigation, misunderstand how the mitigation must be applied, or fail to correctly implement the mitigation in their protocols. Adding further complexity, some of the mitigations can only be applied at the protocol level, and not within  the primitives themselves. 

BLS signatures are also popular as single, unaggregated signatures in blockchain schemes as they are a \emph{unique} signature scheme, meaning for each public key, there is a unique signature per message.
Note that uniqueness is not guaranteed by the EUF-CMA security definition~\cite{lysyanskayaUniqueSignaturesVerifiable2002}.
Although implementors might expect this property to extend to aggregate BLS signatures, it does not, as is demonstrated by the \emph{splitting zero attacks}~\cite{quan2021}, see Section~\ref{sec:background:ZeroSplittingAttack}. 
In fact, it is possible for an attacker to generate malicious public keys and an aggregate signature that will validate for any message, which trivially violates the uniqueness requirement. 
Perhaps surprisingly, splitting zero attacks do not violate the cryptographic security definition of aggregate signatures as no forgery has taken place against an honestly generated key, only against those keys selected by the attacker. At present, there is no mitigation for these attacks in the IETF draft.

Motivated by 
rogue public key attacks and splitting zero attacks, we differentiate between two classes of attacks:
\label{def:behaviors}
\phantom{Hello}
\begin{LaTeXdescription}
	\item[Insecure behaviors:] Attacks on primitives that \emph{violate the EUF-CMA security definition}, e.g., the rogue public key attack.
	\item[Undesirable behaviors:] Attacks on primitives that \emph{satisfy the security definition} but that violate other desirable properties not captured by the security definition, such as the splitting zero attack.
\end{LaTeXdescription}

The above distinction is critical in motivating our different models and understanding the kinds of 
attacks we find, and their scope and impact.
Cryptographic primitives exhibiting insecure behaviors are insecure and should not be used without mitigations.
The situation for undesirable behaviors is less straightforward.
Different applications have different requirements on the primitives and thus an undesirable behavior for one application might be acceptable for another application.
Thus, different primitives with different properties entail different resulting properties for the security protocols and applications using them.

\looseness=-1
Expanding on the above, we note that it is important
to differentiate between attacks on primitives, including insecure and undesirable behaviors thereof, 
and attacks on security protocols.
An adversary could exploit an attack on the primitive to attack a protocol, violating the protocol's security properties.
However, the attack on the primitive alone might not violate the protocol's security properties.
Thus, security protocol analysis must account for the specific properties of the primitives.

With these types of attack in mind, and motivated by the complexity of the protocol-level mitigations that users of aggregate signatures must employ, we turn to building formal models of aggregate signatures suitable for carrying out automated analyses. 
Automated security protocol verification tools, such as ProVerif~\cite{blanchetProVerif03Automatic2021} and the Tamarin prover~\cite{schmidtAutomatedAnalysisDiffieHellman2012}, are very effective in analyzing the properties of security protocols. 
These automated verification tools operate in the `symbolic model' of cryptography, which uses abstractions of the computational definitions to enable automated attack and proof finding.

In this paper, we create the first symbolic models of aggregate signatures.
Modelling multi-party primitives with an arbitrary number of signers, such as aggregate signatures, in the symbolic model is non-trivial.
One challenge is finding a suitable representation of the primitive such that the primitive's properties can be modeled 
without 
blowing up the subsequent analysis.
A second challenge is formulating these properties as an equational theory, as the equations must satisfy certain well-formedness conditions to be used by Tamarin.
Our models support the aggregation of arbitrarily many signatures.

We follow two approaches to modeling aggregate signatures, which results in two classes of models.
Each class consists of three variants accounting for different properties of the primitive, including the behaviors discussed earlier.
Table~\ref{table:AttackFinding:Results}, given 
in Section~\ref{Sec:Evaluation:Comparing}, lists these variants.

Our first class of models is based on the computational security definition of aggregate signatures, assuming only the security 
of signatures that were generated from honestly generated keys.
Thus, this class of models allows for undesirable behaviors such as the splitting zero attack but disallows insecure behaviors such as the rogue public key attack.

To model BLS aggregate signatures specifically, we need to consider 
their behaviors, such as 
the rogue public key attack.
One of the proposed mitigation techniques specified by the IETF draft against the rogue public key attack is enforced on the protocol layer.
This means that protocols using BLS aggregate signatures with this mitigation technique use an insecure primitive that, by itself, is vulnerable to the rogue public key attack, but the protocol should prevent this attack with its own additional checks.
Thus, to verify such protocols, the mitigation technique's effectiveness must be verified for each protocol where it is employed.
We therefore extend our initial model with a specific adversary capability, enabling the rogue public key attack.
As this model allows for all possible undesirable behaviors and even allows the insecure behavior of the rogue public key attack, one can thus prove protocol security properties under very few assumptions, thereby providing strong guarantees when successful.

However, some protocols might rely on aggregate signature primitives
that guarantee additional desirable properties.
Thus, we create a second family of aggregate signature models based on Tamarin's built-in signatures. These models formalize security not only for honestly-generated key pairs, as the computational definition does, but also for all key pairs, honestly generated or not. 
These models reflect desirable behaviors such as uniqueness and thus disallow the splitting zero attack, 
representing primitives with more properties than required by the cryptographic security definitions. 

To close the gap between our first set of models that allows for any undesirable behaviors and our second set of models that excludes all of these behaviors,
we extend the second set of models to 
capture the splitting zero attack and the rogue public key attack. 
This results in a set of models that places stronger restrictions on the attacker's capabilities than our first set of models, and these models 
result in different analysis times,
which we evaluate in Section~\ref{Sec:Eval:Experiments}.

We showcase our aggregate signature models on the remote device attestation protocol SANA~\cite{ambrosinSANASecureScalable2016}, which was published at~CCS'16 with a claimed security proof.
SANA aims to provide scalable and secure remote attestation and uses a novel aggregate signature scheme called optimistic aggregate signatures to reduce verification overhead while minimizing the trust assumptions about the devices and the network.
We modeled
SANA in Tamarin and formalized the security properties stated in~\cite{ambrosinSANASecureScalable2016} and additional authentication properties.
Our analysis revealed necessary assumptions on the device's initialization that were not explicitly stated by SANA's authors.
Without these assumptions, 
Tamarin discovers attacks violating SANA's claimed security properties.
These attacks exploit flaws in the proposed authorization mechanism
and result in 
an incorrect view of the network being presented to the party verifying the attestation.
We verify, using Tamarin, the security properties under the identified assumptions.

\paragraph{Contributions}
We develop the first symbolic models of aggregate
signatures, implemented in Tamarin. Our models are based on the
computational definition of aggregate signatures provided by Boneh et
al.~\cite{bonehAggregateVerifiablyEncrypted2003a} and on BLS aggregate signatures and their IETF
draft~\cite{bonehDraftirtfcfrgblssignature04BLSSignatures}. 
We develop two classes of models: validation models based on the
computational definition of aggregate signatures and attack-finding
models based on Tamarin's built-in signatures. 
Additionally:

\begin{itemize}
  \item We justify the correctness of our models by deriving them from the computational definition of aggregate signatures, which requires choosing appropriate abstractions of the computational definition.
  We address challenges involved with modelling the aggregation of arbitrarily many signatures by introducing different representations of aggregate signatures and techniques such as quantifying over the elements in multisets.
	\item Two attacks have been presented in the literature on BLS aggregate signatures:
	rogue public key attack and splitting zero attack. We formalize those behaviors within our attack-finding models. We show that the validation models allow for splitting zero attacks without explicitly modelling this behavior. In contrast, the rogue public key attack must be modeled explicitly as it contradicts the EUF-CMA security definition. 
	\item We evaluate the behavior of our aggregate signature models on examples, measuring 
	the time required to find proofs and attacks in Tamarin.
		As large numbers of aggregated signatures has a negative impact on Tamarin's run time, we propose simplifications to achieve termination. 
	\item We showcase our models in the security analysis of the remote device attestation protocol SANA~\cite{ambrosinSANASecureScalable2016}.
	We identify necessary assumptions, verify the chosen security properties under these assumptions, and present attacks that are possible without these assumptions and contradict the authors' design requirements.
			
\end{itemize}

\paragraph{Related Work}\label{Sec:Intro:RelatedWork}
Quan~\cite{quan2021,quanAttacksWeaknessesBLS2021} describes several attacks targeting the BLS signature IETF draft v4~\cite{bonehDraftirtfcfrgblssignature04BLSSignatures}.
We focus on the \emph{splitting zero attack}, which 
represents an undesirable behavior of aggregate signatures.
This attack motivates specifying undesirable behaviors in our aggregate signature models.

Jackson et~al.~\cite{jacksonSeemsLegitAutomated2019} points out a gap between the computational definition of signatures and the standard signature model used by all symbolic analysis tools, such as Tamarin and ProVerif, going back to ProVerif's initial publication~\cite{BlanchetCSFW01}. That work only covers `traditional' digital signatures and not aggregate signatures or other types of signatures. 
They point out behaviors of signature implementations that cannot be captured by the standard symbolic model and they define more detailed symbolic models that can capture these behaviors.

In particular, Jackson et~al. define two classes of models: 
The first set of models extends the standard symbolic model of signatures by modelling specific behaviors with explicit additional equations and other techniques.
As one cannot be sure that all possible behaviors lying in the gap between the computational and symbolic model are covered, they introduce a second approach to modeling cryptographic primitives. 
Instead of explicitly allowing certain behaviors, everything that does not contradict the cryptographic definition is allowed.
The cryptographic definition
is formulated using restrictions, which we explain in Section~\ref{Background:Tamarin:Signatures}, restricting the allowed verification results. 
We follow this approach and create models of aggregate signatures using both approaches.

In the same spirit, Cremers et~al.~\cite{cryptoeprint:2024/1920} create abstract models of threshold signatures using Tamarin. 
Their models support an arbitrary number of signers, and capture different security notions 
using different equational theories, restrictions, and adversary capabilities.

Le-Papin et~al.~\cite{le-papinVerifyingListSwarm2023} formalize security requirements for attestation protocols.
They focus on a class of attestation protocols for a large number of devices that let a verifier identify provers with valid and invalid software. 
This class includes the SANA protocol, which we analyze in this work.
Le-Papin~et~al.'s security properties include the prover's authentication to the verifier, which we also consider in our analysis of SANA.
They focus on the SIMPLE+ protocol and model it with Tamarin.
For this analysis they use our aggregate signature models from 
the unpublished master thesis~\cite{hofmeierFormalizingAggregateSignatures2021}.
They chose the attack-finding models, as they are faster.

\paragraph{Outline}
In Section~\ref{Sec:Background}, we provide background on aggregate signatures and the Tamarin prover.
In Section~\ref{Sec:ValidationModel}, we present our validation models, which follow the computational definition of aggregate signatures and we extend them for the rogue public key attack.
In Section~\ref{Sec:AttackFindingModels}, we present our attack-finding models, based on the standard signature model and extensions for colliding signatures and rogue public key attacks.
We evaluate these models in Section~\ref{Sec:Evaluation} by comparing them on a simple protocol and comparing their proof times.
In Section~\ref{Sec:SANA}, we showcase our models on a security analysis of SANA and, in Section~\ref{Sec:Conclusion}, we draw conclusions.

\section{Background}\label{Sec:Background}
In this section, we provide background on aggregate signatures and the Tamarin prover.

\subsection{Aggregate Signatures}\label{Sec:Background:AggSign}
Aggregate signature schemes~\cite{bonehAggregateVerifiablyEncrypted2003a} compress multiple signatures $\bm{\sigma} = (\sigma_1, . . . , \sigma_n)$ of different messages into one short aggregate signature $\sigma_\text{agg}$. 
This compression reduces storage and bandwidth requirements, and is useful in situations where many independent signers interact.

An \emph{aggregate signature scheme} is a signature scheme with the two additional algorithms, $\aggComp$ and $\vfyAggComp$.
The \emph{aggregation algorithm}'s $\aggComp$ input is a vector of public keys $\mathbf{pk} = (pk_1, . . . , pk_n )$ and a vector of signatures $\bm{\sigma} = (\sigma_1, . . . , \sigma_n)$ and it outputs an aggregate signature $\sigma_\text{agg}$.
The deterministic \emph{aggregate verification algorithm}'s $\vfyAggComp$ input is a vector of public keys $\mathbf{pk} = (pk_1 , . . . , pk_n)$, a vector of messages $\mathbf{m} = (m_1, . . . , m_n)$, and an aggregate signature $\sigma_\text{agg}$. 
It outputs either $\true$ or $\false$.
The scheme is \emph{correct} if for all $\mathbf{pk}$,
$\mathbf{m}$,
and $\bm{\sigma}$,
if $\vfyComp(pk_i, m_i , \sigma_i) = \true$ for $1 \leq i \leq n$, then 
$Pr[\vfyAggComp(\aggComp(\mathbf{pk}, \bm{\sigma}), \mathbf{m}, \mathbf{pk}) = \true] = 1$.
We provide a more formal definition in Appendix~\ref{Appendix:AggDefinitions}, Definition~\ref{def:background:signatureAggregation}.
Signatures can be aggregated without knowledge of the signing keys. 
Thus, any agent can aggregate the signatures that it possesses.

Aggregate signatures have a well-established security definition~\cite{bonehGraduateCourseApplied}, which is the natural extension of \emph{\ac{EUF-CMA}} to the many-signer case.
Here, we describe the attack game and we provide the detailed definition in Appendix~\ref{Appendix:AggDefinitions}, Definition~\ref{def:background:attackGameAggregateSignature}.

The challenger generates a key pair 
$(pk, sk) \leftarrow \genComp()$
and sends $pk$ to the adversary.
The adversary sends the challenger signing queries, each consisting of a message $m^{(i)}$.
The challenger responds to each signing query with 
$\sigma^{(i)} \leftarrow \signComp(m^{(i)}, sk)$.
Eventually, the adversary outputs a candidate aggregate forgery $(\mathbf{pk}, \mathbf{m}, \sigma_\text{agg})$, where $\mathbf{pk} = (pk_1 , . . . , pk_n )$ and $\mathbf{m} = (m_1, . . . , m_n)$.
The adversary wins the game if $\vfyAggComp(\sigma_\text{agg}, \mathbf{m}, \mathbf{pk}) = \true$ and one of the public keys $pk_j \in \mathbf{pk}$ is the public key $pk$ generated by the challenger where the adversary did not issue a signing query for the corresponding message $m_j$.

\subsubsection{BLS Aggregate Signatures}\label{Sec:Background:BLS}

Boneh–Lynn–Shacham (BLS) aggregate signatures~\cite{bonehAggregateVerifiablyEncrypted2003a} were the first instantiation of aggregate signatures.
These signatures are used in various applications, resulting in their standardization in an IETF draft~\cite{bonehDraftirtfcfrgblssignature05BLSSignatures2022}.
We provide a full description of BLS signatures and their properties 
\paper{in the extended version of this paper~\cite{extendedVersion},}\extended{in Appendix~\ref{Appendix:BLS},}
but introduce their key properties below.

\label{sec:background:RogueKeyAttack}
Naive constructions of \acs{BLS} aggregate signatures are vulnerable to the \emph{rogue public key attack}. 
This enables an attacker, through a malicious choice of their public key, to forge an aggregate signature that verifies under the victim's and the attacker's public keys.
\paper{See the extended version of this paper~\cite{extendedVersion} for details.}\extended{See Appendix~\ref{Appendix:BLS}, Section~\ref{appendix:Background:RoguePublicKeyAttack}, for details.}
The IETF draft~\cite{bonehDraftirtfcfrgblssignature05BLSSignatures2022} defines three mitigations to prevent this:
\begin{enumerate}
	\item Reject duplicate messages during the verification. 
	\item Bind signed messages to their public keys.
	\item Require signers to prove possession of their secret key.
\end{enumerate} 
These techniques 
have different trade-offs
and are thus suited for different applications.
The first two techniques are part of the verification and signing algorithm and thus they are part of the primitive.
However, the third technique requires additional checks to be carried out by the protocol.

In contrast to the aggregate signature definition, Appendix~\ref{Appendix:AggDefinitions}, Definition~\ref{def:background:signatureAggregation}, the aggregation of \acs{BLS} signatures does not require the public keys.
As the aggregation does not require one to validate the aggregated signatures, we also omit the public keys in our models.

\subsection{The Tamarin Prover}\label{Sec:Background:Tamarin}
\looseness=-1
The Tamarin Prover \cite{meierTAMARINProverSymbolic2013,
schmidtAutomatedAnalysisDiffieHellman2012} is an automated
verification tool used to model and analyze security protocols.
Given a protocol model and a security property,
Tamarin either constructs a proof or finds an attack on the
property. As the underlying problem is undecidable, the tool
may not always terminate.

Tamarin operates in the symbolic model of cryptography where cryptographic messages are
represented as terms. Each cryptographic message is a constant, a
variable, or a function symbol applied to some messages. The function
symbols represent cryptographic primitives or grouping
operations. The properties of these functions are expressed with
equational theories. 
The equations model, for example, that decrypting an
encrypted ciphertext, using the appropriate decryption key, yields the original plaintext, as expected.
Tamarin supports various built-in equational theories for
cryptographic primitives, such as for signing, which we introduce shortly.
Tamarin also supports user-defined function symbols and equations,
which must satisfy certain well-formedness conditions \cite{tamarinProverManual}.

Tamarin works with a model of a standard Dolev-Yao adversary that
controls the network. Thus, the adversary learns any messages sent
over the network and can derive messages from its knowledge.
However, cryptography is assumed to work perfectly.
This means, for example, that for decryption the adversary needs
the appropriate decryption key, and importantly it has no cryptanalytic
capabilities.
Hence, given a cipher text, the adversary learns the 
plaintext if and only if it possesses the associated decryption key.
The adversary can also apply functions to messages in its knowledge, for example encrypting messages with keys it knows.

Tamarin models are specified using terms, facts, and rules.
We already presented the terms that represent messages.
Facts are of the form \texttt{F(t1,...,tk)}, where \texttt{F} is a
fact symbol with some fixed arity and \texttt{ti} are
terms. 
Fact symbols are used to represent protocol participant
states,
messages on the network, the adversary's knowledge, and the generation of fresh nonces.
Fact names are user-chosen and have no intrinsic meaning, with the exception of some special built-in names we present next.
Sending a message \texttt{m} to the network is represented by the 
fact \texttt{Out(m)} and receiving a message \texttt{m} is represented
by the fact \texttt{In(m)}. The generation of a fresh
nonce \texttt{\nicetilde n} is represented by \texttt{Fr(\nicetilde
 n)}.

A labeled multiset rewrite rule consists of the keyword \texttt{rule} followed by the rule name and three multisets of facts: the premises, the actions (also called labels), and the conclusions.
The following rule, called \texttt{Example}, has its premise on line 2, action on line~3, and conclusions on line~4.
\begin{lstlisting}[language=Tamarin, escapechar=*]
rule Example:
  [ Fr(~m) ] *\label{line:background:exampleRule:left}*
--[ Label(~m) ]-> *\label{line:background:exampleRule:lable}*
  [ State(~m), Out(~m) ] *\label{line:background:exampleRule:right}*
\end{lstlisting}
In this rule, an agent creates a fresh nonce \texttt{\nicetilde m}, saves it in the agent's state and sends this nonce
to the network. The rule is labeled with the action fact \texttt{Label(\nicetilde m)}.

Tamarin considers arbitrarily many protocol executions interleaved in parallel.
These executions,
also capturing the adversary's behavior, 
are represented
as a labeled transition system. 
The protocol's state is represented by a multiset of facts representing the local states of the protocol participants,
the messages on the network, and the adversary's knowledge. 
Labeled multiset rewrite rules model both the protocol steps and the adversary capabilities. 
The protocol can transition into a new state by applying such a
rule to the protocol state. 
A rule can be applied if there is a subset of facts in the current
state that matches the rule's premises. Applying the rule removes
this matched subset of facts from the state and adds instantiations of
the rule's conclusions under the matching substitution. 

A protocol's execution is represented by the repeated application of multiset rewrite rules to the protocol's state, where the initial state is the empty multiset. 
The trace of such an execution is defined by the sequence of the instantiated action facts of the rewrite rules.

The semantics of a security protocol is defined as the set of all traces of the protocol's labeled transition system. 
We express the protocol's security properties as trace properties, which are also sets of traces. 
Hence, a protocol satisfies a security property if the protocol's set of traces is a subset of the property's traces. 
If this is not the case, counter examples (i.e., attacks) exist that violate the property. 

In Tamarin, we formulate security properties as first-order
formulas, called \emph{lemmas}. There are two kinds of lemmas: the most
common one expresses that the property must hold for all protocol traces. Alternatively, \emph{executability lemmas}, marked with
the keyword \texttt{exists-trace}, state that there exists a trace for
which the property holds. The first kind of lemma states
that a property holds for all protocol behaviors. The second kind is
mostly used to check that the protocol model is executable, increasing
one's confidence in the given model.

The set of traces considered by Tamarin can be restricted using
\emph{restrictions},
which are first-order formulas, with a syntax
similar to lemmas. 
Restrictions express properties that the traces considered by Tamarin must
have. 
They can be used, for example, to enforce that certain rules
are only applied once, or 
to formalize signature verification.

\subsubsection{Signatures in Tamarin}\label{Background:Tamarin:Signatures}
Tamarin provides the following built-in equational theory for signatures.
\begin{lstlisting}[language=Tamarin, label={listing:RuleBased:SignatureModel}]
  functions: sign/2, verify/3, pk/1, true/0
  equations: verify(sign(m,sk), m, pk(sk)) = true
\end{lstlisting}
This theory provides four function symbols for signing, verification, deriving a public key from a secret key, and \texttt{true}, which represents the verification result.
The equation expresses that signatures verify (they equal true) exactly if the message and public key match the signed message and signing key.

This is a coarse abstraction of the computational definition
as noted in \cite{jacksonSeemsLegitAutomated2019}.
There, the correctness property gives guarantees only for honestly-generated key pairs and the \ac{EUF-CMA} security definition guarantees that all efficient adversaries have a negligible advantage of winning the attack game for honestly-generated keys.
The computational definition provides no guarantees for signatures with not honestly-generated keys.
Prior symbolic models guarantee correctness and security even for not honestly-generated keys.
We now refer to honestly-generated keys and signatures using these keys as \emph{honest keys} and \emph{honest signatures}.
Keys and signatures that are not honestly generated are called \emph{non-honest keys} and \emph{signatures}.

\section{Validation Models}\label{Sec:ValidationModel}

\looseness=-1
In this section, we describe our validation models and in the next section we describe our attack-finding models.
All models are available at~\cite{hofmeierTamarinModelsAggregate}.
Moreover, for additional details, see~\cite{hofmeierFormalizingAggregateSignatures2021}.

Our validation models are based on the 
computational definition of aggregate signatures.
Here, we represent the verification of an aggregate signature $\sigma_\text{agg}$ using predicate notation: $\vfyAgg(\sigma_\text{agg}, \mathbf{m}, \mathbf{pk}, b)$, where $b$ is the verification result.
In these models, every behavior that is not explicitly forbidden by the definition is possible.
Hence each 
verification $\vfyAgg(\sigma_\text{agg}, \mathbf{m}, \mathbf{pk}, b)$ functions as an adversary controlled oracle whose
verification result $b$ 
could be either $\true$ or $\false$\footnote{For our validation models, we introduce the nullary function \texttt{false}.} unless governed by a restriction.

Our validation models rely on the computational definition of aggregate signatures,
namely on their correctness, consistency, and on \ac{EUF-CMA} security or unforgeability.
We say that a signature or aggregate signature \emph{verifies} if the verification algorithm outputs $\true$.
The \emph{correctness} of aggregate signatures states that if all signatures $\sigma_i$ in an aggregation $\aggComp(\sigma_1, ..., \sigma_n)$ verify, the aggregation must verify as well. 
\emph{Unforgeability} states that if an aggregate signature verifies, 
all aggregated signatures $\sigma_i$ either verify or the public key $pk_i$ is not honest.
\emph{Consistency} states that the verification algorithm $\vfyAggComp$ is deterministic, meaning a given verification must always return the same result. 
We model the verification of aggregate signatures with these three properties: correctness, unforgeability, and consistency. 
Next, we formalize these properties as restrictions.

\subsection{Formalizing the Properties}\label{Sec:VerifModels:Restrictions}

The correctness definition of aggregate signatures, Appendix~\ref{Appendix:AggDefinitions}, Definition~\ref{def:background:signatureAggregation}, states that if all signatures in an aggregation verify, the aggregation must verify as well.
We combine this definition with the correctness definition of signatures stating that for an honestly-generated key pair $(sk, pk)$, the signature $\sigma = \sign(m, sk)$ must verify for message $m$ and public key $pk$.
This results in the following restriction, where $\Honest(pk)$ states
that the public key $pk$ was honestly generated and $sk_j$ is the secret key corresponding to the public key $pk_j$.
\begin{tamarinrestriction}[Correctness of aggregate signatures]\label{restr:restrictionBased:Correctness}\label{restr:Correctness}
	\begin{multline*}
	\forall \mathbf{pk}, \mathbf{m}, \bm{\sigma}. (\forall i \in \{1, ..., n\}. (\Honest(pk_i) \land\\ \sigma_i = \sign(m_i, sk_i)) \Rightarrow \vfyAgg(\agg(\bm{\sigma}), \mathbf{m}, \mathbf{pk}, \true))
	\end{multline*}
\end{tamarinrestriction}

The \ac{EUF-CMA} security definition of aggregate signatures, 
Appendix~\ref{Appendix:AggDefinitions}, 
Definition~\ref{def:background:attackGameAggregateSignature}, 
states that
no adversary can win 
the attack game with a non-negligible probability. 
We abstract this definition for the symbolic model and assume that no adversary can win the attack game.
Winning the attack game means, after having access to a signing oracle, the adversary produces a forgery $(\mathbf{pk}, \mathbf{m}, \sigma_\text{agg})$ where $\vfyAgg(\sigma_\text{agg}, \mathbf{m}, \mathbf{pk}, \true)$, where one public key $pk_j \in \mathbf{pk}$ is honestly generated,
the corresponding secret key $sk_j$ is not known to the adversary, and the corresponding message $m_j \in \mathbf{m}$ was not queried.
As we assume that such a forgery is not possible, for each triple $(\mathbf{pk}, \mathbf{m}, \sigma_\text{agg})$, 
each public key $pk_i \in \mathbf{pk}$ is either not honest or there is a corresponding signature $\sigma_i$ that is honestly created, which means that $\sigma_i = \sign(m_i, sk_i)$. 
Note that $\sigma_{\text{agg}}$ can be an arbitrary term if all public keys are not honest. 
Only if some public keys are honest, must it be an aggregation of signatures $\sigma_{\text{agg}} = \agg(\bm{\sigma})$. 
We express this as the following property:
\begin{multline}\label{eq:restriction:security:complicated}
  \forall \mathbf{pk}, \mathbf{m}, \sigma_\text{agg}. \vfyAgg( \sigma_\text{agg}, \mathbf{m}, \mathbf{pk}, \true)
  \\ \Rightarrow (\forall i \in \{1, ..., n\}. \lnot \Honest(pk_i) \\
  \lor (\exists \bm{\sigma}. \sigma_\text{agg} = \agg(\bm{\sigma}) 
  \land (\forall i \in \{1, ..., n\}. \neg \Honest(pk_i)  \\ \lor \exists \sigma_j \in \bm{\sigma}. \sigma_j = \sign(m_i, sk_i) ) )) \text{.}
\end{multline}
To improve Tamarin's termination, we simplify this restriction by assuming the following:

\begin{itemize}
	\item The message and public key vectors $\mathbf{m}$ and $\mathbf{pk}$ have the same length.
	\item The signature aggregation $\sigma_\text{agg}$ is of the form $\agg(\bm{\sigma})$.
	\item There must be a signature $\sigma_i$ for each public key and message pair $(pk_i, m_i)$, and vice versa.
	    \end{itemize}

We justify these assumptions with input validations that are performed in practice.
For example, the IETF draft for BLS signatures~\cite{bonehDraftirtfcfrgblssignature05BLSSignatures2022}
requires the public key and message vectors to be of the same length.
We enforce the above assumptions with some additional restrictions.
With these assumptions, we can simplify Property~\ref{eq:restriction:security:complicated} and get the following unforgeability restriction.

\begin{tamarinrestriction}[Unforgeability of aggregate signatures]\label{restr:Security}
  \begin{multline*}
  \forall \mathbf{pk}, \mathbf{m}, \bm{\sigma}. \vfyAgg(\agg(\bm{\sigma}), \mathbf{m}, \mathbf{pk}, \true) \\\Rightarrow (\forall i \in \{1, ..., n\}. (\neg \Honest(pk_i) \lor \sigma_i = \sign(m_i, sk_i)))
  \end{multline*}
\end{tamarinrestriction}

The verification algorithm's deterministic behavior is expressed as the following consistency restriction. 
\begin{tamarinrestriction}[Consistency of aggregate signatures]\label{restr:Consistency}
	\begin{align*}
	\forall \mathbf{pk}, \mathbf{m}, \sigma_\text{agg}, b_1, b_2.
	&\vfyAgg(\sigma_\text{agg}, \mathbf{m}, \mathbf{pk},  b_1) \nonumber\\\land
	&\vfyAgg(\sigma_\text{agg}, \mathbf{m}, \mathbf{pk}, b_2 )
	\Rightarrow b_1 = b_2
	\end{align*}
\end{tamarinrestriction}

Usually, we only model honest keys in Tamarin as standard primitives and do not differentiate between honest and non-honest keys.
However in our restriction-based models, correctness and unforgeability restrictions only specify the verification results for honest keys.
Signature aggregations of non-honest keys can verify arbitrarily, while fulfilling the consistency restriction.
We thus add registration rules of non-honest keys for the adversary.
This enables various undesirable behaviors, such as the splitting zero attack, which we will discuss in Section~\ref{sec:background:ZeroSplittingAttack}.
In Section~\ref{Sec:Evaluation:Comparing}, we will compare our validation model without non-honest keys, which does not allow undesirable behaviors with non-honest keys, to the validation model with non-honest keys, which allows such behaviors.

\subsection{Signature Aggregation in Tamarin}\label{Sec:ValidationModel:AggInTamarin}

We represent signature aggregations by the function \texttt{agg/1} applied to a multiset of signatures.
As the correctness and unforgeability restriction require accessing signatures, messages, and public keys of the same index, we introduce explicit indices.
For example, we model $\agg(\sign(m_1, pk_1), \sign(m_2, pk_2))$ as
\begin{lstlisting}[language=Tamarin,numbers=none,escapechar=*]
agg(<sign(m1,pk1), ind_1> + <sign(m2, pk2), ind_2>)*$,$*
\end{lstlisting}
\noindent where multisets are represented by the associative-com\-mu\-ta\-ti\-ve operator~\texttt{+} and tuples containing the terms \texttt{a} and \texttt{b} are represented by \texttt{<a,b>}.
The indices \texttt{ind\_1}, \ldots, \texttt{ind\_n} can be modeled in Tamarin in various ways, such as fresh values, constants, counters, or values provided by the adversary.
Each option has different trade-offs. 
We opted for fresh values as they are always distinct.
However, they must be provided to the adversary after aggregation, as indices should be public.

To model the validation of a signature aggregation, the action fact \texttt{VfyAgg} is added to the appropriate agent rule.
The arguments are the signature aggregation, a multiset of triples of message, public key, and index, and the expected verification result.
For example, the validation of the signature aggregation \texttt{agg} on messages \texttt{m1}, \texttt{m2} and public keys \texttt{pk1}, \texttt{pk2}, with the result $\true$, is represented by the following action fact:
\begin{lstlisting}[language=Tamarin,numbers=none,escapechar=*]
VfyAgg(agg, <m1,pk1,ind_1> + <m2,pk2,ind_2>, true)*$.$*
\end{lstlisting}

\looseness=-1
The abstract restrictions described in Section~\ref{Sec:VerifModels:Restrictions} must be adapted for Tamarin. 
This includes case distinctions for one or multiple aggregated signatures and translating the quantification over the indices into Tamarin's syntax.
The latter utilizes the associative-commutative property of Tamarin's multisets.
Namely, the quantification over all pairs in a multiset can be expressed as \texttt{All a b r. <a, b> + r}.
Note that this quantification requires the multiset's elements to be identifiable by an applied function such as a pair.
As otherwise pattern matching for an arbitrary term \texttt{x} in \texttt{x~+~y},
both \texttt{x} and \texttt{y} could be single elements or multisets themselves.
See Appendix~\ref{Appendix:TamarinRestrictions}
for more details on formalizing the restrictions in Tamarin.

\subsection{Rogue Public Key Attack}\label{sec:ValidationModel:RoguePublicKey}

The validation models discussed so far are based on the  EUF-CMA security definition of aggregate signatures and include no further desirable properties.
Thus, as mentioned, the validation model with non-honest key registration permits undesirable behaviors that conform with our restrictions, such as the splitting zero attack.
However, the rogue public key attack, introduced in Section~\ref{sec:background:RogueKeyAttack},
violates the security definition and is thus not permitted by this model.
As mentioned, one of the mitigation techniques against the rogue public key attack, Proof of Pos\-ses\-sion, relies on the protocol layer.
To verify the effectiveness of such protocol-based mechanisms,
we require models of aggregate signatures that are vulnerable to the rogue public key attack.
Thus, we extend our validation models to model BLS aggregate signatures with a protocol-based mitigation by adding an adversary capability for a rogue public key attack.

First, we provide a high-level description of the rogue public key attack and then present our Tamarin model extension.
\paper{See the extended version of this paper~\cite{extendedVersion} for a detailed description of this attack.}\extended{See Appendix~\ref{Appendix:BLS}, Section~\ref{appendix:Background:RoguePublicKeyAttack} for a detailed description of the rogue public key attack.}

The rogue public key attack allows an adversary to 
calculate a rogue public key $pk_\text{rogue}$ from a target public key $pk_\text{target}$ without being able to calculate the corresponding rogue secret key.
However, the adversary can create a rogue signature aggregation $\sigma_{\text{agg}_\text{rogue}}$ that is valid for a chosen message $m$ repeated twice, the target public key $pk_\text{target}$ and the rogue public key $pk_\text{rogue}$:
\begin{equation*}
  \vfyAggComp(\sigma_{\text{agg}_\text{rogue}}, (m, m), (pk_\text{target}, pk_\text{rogue})) = \true \text{.}
\end{equation*}
Thus, the adversary creates a forgery for the target public key.

We enable the adversary to create rogue signature aggregations with additional adversary rules, see~\cite{hofmeierFormalizingAggregateSignatures2021} for details.
The first rule creates a rogue public key from a target public key and the other rules create a rogue signature aggregation from a rogue public key. 
The rogue signature aggregation is represented as a regular signature ag\-gre\-ga\-tion that can be verified as any other signature aggregation.
Thus, the aggregation contains a forged signature of the target key and a signature of the rogue key.
This second signature therefore contains the rogue secret key.
Note, in practice, that the adversary creates the rogue public key without the secret key. 
Thus in our model, the adversary must not learn the rogue secret key.
We therefore use special \emph{private} functions that the adversary cannot apply, 
which can only be used in rules.
We introduce the private function \texttt{rogueSk/1} to represent the rogue secret key of some target public key.

The first added adversary rule provides the rogue public key \\\texttt{pk(rogueSk(pk\_target))} for a given target key \texttt{pk\_target} and registers the rogue public key in the \ac{PKI}.
The second added rule provides the rogue signature aggregation for a rogue public key \texttt{pk(rogueSk(pk(skTarget)))} and a message \texttt{m}. 
Note that it does not require the knowledge of the honest signature \texttt{sign(m, skTarget)}.
The returned rogue signature aggregation has the form
\begin{lstlisting}[language=Tamarin,numbers=none,escapechar=*]
agg(<sign(m, skTarget), index_target> 
  + <sign(m, rogueSk(pk(skTarget))), index_rogue>)*$.$*
\end{lstlisting}
\noindent Note that this rule extracts the target secret key \texttt{skTarget} and the rogue secret key \texttt{rogueSk(pk(skTarget))} to create the rogue signature aggregation, while the adversary cannot access either of the two secret keys.
A third adversary rule
aggregates additional signatures to the rogue aggregate.

\section{Attack-Finding Models}\label{Sec:AttackFindingModels}

The validation models presented so far only guarantee unforgeability, correctness, and consistency, but no further desirable properties (see Section~\ref{def:behaviors}).
As the unforgeability and correctness restrictions only define the behavior of aggregate signatures with honestly-generated keys, the behavior of non-honest keys can be arbitrary and indeed these models allow for many undesirable behaviors with non-honest keys.
This supports proving security properties of protocols under very few assumptions, which provides strong guarantees.
However, some security protocols might rely on primitives with stronger guarantees.
Our second family of models, the attack-finding models described in this section, forbids any undesirable behaviors with non-honest keys by guaranteeing unforgeability and correctness not only for honestly-generated key pairs, but for all key pairs.

Our attack-finding models of aggregate signatures are based on Tamarin's built-in signature model, described in Section~\ref{Background:Tamarin:Signatures}.
We define 
\begin{equation*}
  \vfyAgg(\sigma_\text{agg}, (m_1, ..., m_n),(\pk(sk_1), ...,\pk(sk_n))) = \true
\end{equation*}
if and only if $\sigma_\text{agg} = \agg(\sigma_1, ..., \sigma_n)$, where $\sigma_i =\sign(m_i, sk_i)$ for all $1 \leq i \leq n$.
Note that this definition considers all key pairs, honest or not.
This is an abstraction of the EUF-CMA security definition of aggregate signatures,
Appendix~\ref{Appendix:AggDefinitions},
Definition~\ref{def:Background:BLS:security}, which only considers honestly-generated keys.

A naive approach to defining $\vfyAgg$ with recursive equations fails 
due to Tamarin's requirements on user-defined equations.
Thus, we model the verification with a single, non-recursive equation that checks that the messages and keys of the aggregated signatures match the messages and keys provided for the verification.
\begin{lstlisting}[language=Tamarin]
equations: 
vfyAgg(agg(m_list, pk_list), m_list, pk_list) = true 
\end{lstlisting}
We represent an aggregate signature as the lists of messages \texttt{m\_list} and public keys \texttt{pk\_list} of the aggregated signatures.
These lists can be modeled using Tamarin's built-in tuples or multisets.
We choose tuples to more closely resemble the vectors from the computational definition.
We represent a signature aggregation $\agg(\sign(m_1, sk_1), ..., \sign(m_n, sk_n))$ by
\begin{lstlisting}[language=Tamarin,numbers=none,escapechar=*]
agg(<m1, ..., mn>, <pk(sk1),..., pk(skn)>)*$.$*
\end{lstlisting}
Note that in Tamarin's symbolic model of signatures, public keys can be directly derived from secret keys.
Thus representing an aggregate signature using public keys only requires the application of the \texttt{pk} function.
Using public keys instead of secret keys allows for direct comparison in the above equation.

Agents aggregating signatures must extract the signed messages and signing keys from the signatures.
The following example shows a rule where an agent aggregates the two signatures \texttt{sign(m1, sk1)} and \texttt{sign(m2, sk2)}.

\begin{lstlisting}[language=Tamarin]
rule Aggregator_aggregates_two_signatures:
  [ A_1(sign(m1, sk1), sign(m2, sk2)) ] -->
  [ A_2(agg(<m1, m2>, <pk(sk1), pk(sk2)>)) ]
\end{lstlisting} 

\noindent Note that this rule implicitly verifies the signatures using pattern matching.

To prevent the adversary from directly creating an aggregate signature from a list of public keys and messages without knowing the signatures, the function \texttt{agg} is private.
Thus, the adversary cannot apply \texttt{agg}, but 
agents aggregating signatures can be modeled as in the above rule.

In practice, anyone can aggregate signatures, also the adversary.
Thus, we 
provide adversary rules for incremental signature aggregation.
Note that not all aggregate signature schemes offer incremental aggregation and our models could be adapted for non-incremental aggregation.
Also note that because Tamarin's tuples are not associative, 
the adversary aggregation rules and the agent aggregation rules must produce and accept aggregations of the same shape, namely listing the messages and public keys as \texttt{<a1, <a2,..,<an-1,an>..>} and not as \texttt{<..<a1,a2>,..,an-1>,an>}.

The adversary aggregation rules only allow the aggregation of valid signatures. 
Thus, the resulting aggregate signature is valid by construction. 
An invalid aggregation can be modeled by an arbitrary \texttt{term}. 
The verification of this arbitrary term will not validate, as the 
verification equation cannot be applied to \texttt{vfyAgg(term, m, k)}.

So far, we created an abstract model of aggregate signatures
which behaves `perfectly'.
In the next two subsections, we extend this model to model BLS signatures by modeling 
their real-world behavior and known attacks.

\subsection{Rogue Public Key Attack}\label{AttackModels:RogueAttack}

As discussed in Section~\ref{sec:ValidationModel:RoguePublicKey}, some versions of BLS aggregate signatures mitigate the rogue public key attack on the protocol layer.
We thus add an adversary capability for the rogue public key attack.

We extend our model by adding an adversary rule, similar to the validation model with the rogue public key attack described in Section~\ref{sec:ValidationModel:RoguePublicKey}.
In contrast to the validation model extension, we do not model the rogue secret key but only the rogue public key.
We model rogue public keys using the function \texttt{roguePk/1} applied to the target public key. 
The adversary can register these rogue public keys in the \ac{PKI} with an additional rule.

We differentiate between honest signature aggregations and rogue signature aggregations by renaming the function \texttt{agg/2} to \texttt{validAgg/2} and introducing the new (also private) function \texttt{rogueAgg/2}.
Differentiating between valid and rogue signature aggregations makes reasoning about attacks easier, but it is not necessary.
Both aggregations could also be modeled using one function symbol.
The verification of valid signature aggregations and rogue signature aggregations behave the same.
This is modeled by an additional equation for the verification of rogue signature aggregations. 
The adversary creates rogue signature aggregations using an additional aggregation rule.
\begin{lstlisting}[language=Tamarin]
rule Adv_RogueKey_Aggregation_new:
[In(<m, roguePk(pkTarget)>)] -->
[Out(rogueAgg(<m,m>, <pkTarget,roguePk(pkTarget)>))]
\end{lstlisting}
\noindent
This rule creates a 
rogue signature aggregation for a target public key and a corresponding rogue public key.
Additional rules allow the adversary to further aggregate combinations of rogue signature aggregations, valid signature aggregations, and valid signatures.

\subsection{Colliding Signatures}\label{Sec:AttackFindingModel:CollidingSignature}
Our standard attack-finding model guarantees unforgeability and correctness for all keys, honestly generated or not.
Thus, this model guarantees desirable properties such as uniqueness, which states that for each vector of public keys, there is a unique aggregate signature for every message vector.
Uniqueness holds for single BLS signatures.
However, uniqueness fails for BLS \emph{aggregate} signatures, as demonstrated by the splitting zero attack. 
To more closely model BLS aggregate signatures, we extend this model to capture this attack.

We first describe this attack at a high level.
We then provide a more general definition of colliding signatures
that includes the splitting zero attack and we extend our model for colliding signatures.
\paper{See the extended version of this paper~\cite{extendedVersion} for more details on BLS aggregate signatures and the splitting zero attack. }\extended{See Appendix~\ref{Appendix:BLS} for more details on BLS aggregate signatures and the splitting zero attack. }

\subsubsection{Splitting Zero Attack}\label{sec:background:ZeroSplittingAttack}
Quan~\cite{quan2021,quanAttacksWeaknessesBLS2021} describes several attacks on BLS aggregate signatures, with a focus on the IETF draft~v4~\cite{bonehDraftirtfcfrgblssignature04BLSSignatures}.
We focus here on the splitting zero attack.

Quan observed that for single BLS signatures,
any signature 
signed with a non-honest key $sk_\text{adv} = 0$ verifies for any message.
As the corresponding public key 
is the identity element, this attack can be easily detected, as done by the BLS IETF draft~\cite{bonehDraftirtfcfrgblssignature04BLSSignatures}, with a check in the verification algorithm.

Quan~\cite{quan2021} extends this attack for BLS aggregate signatures and calls it the \emph{splitting zero attack}.
The attack bypasses the identity element check by using multiple non-honest keys.
The adversary chooses two non-honest keys, such that $sk_{\text{mal}_1} + sk_{\text{mal}_2} = 0$ and $sk_{\text{mal}_1}$ and $sk_{\text{mal}_2} $ are each non-zero. 
Thus, the corresponding public keys 
will validate. 
However, aggregating non-honest signatures using these non-honest keys $\signComp(m, sk_{\text{mal}_i})$ and honest signatures $\bm{\sigma}$ results in an aggregate signature that verifies for different message vectors.
Namely, the aggregate signature $\aggComp(\signComp(m, sk_{\text{mal}_1}), \signComp(m, sk_{\text{mal}_2}), \bm{\sigma})$ validates against the message vector $(m', m', \mathbf{m})$, where $\mathbf{m}$ is the message vector signed by $\bm{\sigma}$ and $m'$ can be any message and not the signed message $m$ as expected.

This attack does not violate the EUF-CMA security definition of aggregate signatures provided in 
Appendix~\ref{Appendix:AggDefinitions},
Definition~\ref{def:Background:BLS:security}, as it creates a non-honest signature for a non-honest key and not a signature forgery for an honest key. 
It can still be an unexpected and undesirable behavior for a protocol designer. 

\subsubsection{Extending the model}

We generalize the splitting zero attack as colliding signatures.
We define colliding signatures of aggregate signatures, analogous to colliding signatures of single signatures according to Jackson et~al.~\cite{jacksonSeemsLegitAutomated2019}, as follows:
For non-honest keys $sk_{\text{mal}_1}, ..., sk_{\text{mal}_k}$, messages $m_1, ..., m_k,$ $m_l, ..., m_n$, honest keys $sk_l, ..., sk_n$, and honest signatures $\sigma_l = \sign(m_l, sk_l), ..., \sigma_n=\sign(m_n, sk_n)$, the aggregation 
\begin{equation}\label{equation:collidingAggregation}
\sigma_{\text{agg}_\text{mal}} = \agg(\sigma_{\text{mal}_1}, ..., \sigma_{\text{mal}_k},\sigma_l, ... ,\sigma_n),
\end{equation}
where ${\sigma_{\text{mal}_i} = \sign(m_i, sk_{\text{mal}_i})}$,
will validate against any messages $m_1', ..., m_k'$ and the messages $m_l, ..., m_n$:
\begin{multline}
\vfyAgg(\sigma_{\text{agg}_\text{mal}}, (\textcolor{red}{\bm{m_1', ..., m_k'}}, m_l, ..., m_n),\\ (pk_{\text{mal}_1}, ..., pk_{\text{mal}_k}, \pk(sk_l), ..., \pk(sk_n))) = \true.
\end{multline}
\noindent
With these colliding signatures, we can capture the splitting zero attack, which is a special case of a colliding signature where $m_1 = m_2 = ... = m_k$ and $m_1'= m_2' = ... = m_k'$.

We model colliding signature aggregations with the private function \texttt{zeroAgg/2}.
The first argument is a valid signature aggregation and the second is a list of non-honest public keys.
The colliding signature aggregation from Equation~\ref{equation:collidingAggregation} is modeled as:
\begin{lstlisting}[language=Tamarin,numbers=none,escapechar=*]
zeroAgg(validAgg(<ml,...,mn>,<pk(skl),...,pk(skn)>)
            , <pk(skMal1), ..., pk(skMalk)>)*$.$*
\end{lstlisting}

The verification equations only check the messages and public keys of the honest signatures, and the non-honest public keys are ignored.
Each verification equation covers a fixed number of non-honest signatures.
Thus, our model cannot cover arbitrarily many non-honest signatures.
The following equation, for example, models the case of one non-honest key. 
\begin{lstlisting}[language=Tamarin]
equations: vfyAgg(zeroAgg(validAgg(m, k), kZero)
             , <m_non_honest, m>, <kZero, k>) = true 
\end{lstlisting}

The registration of non-honest keys and the aggregation of colliding signatures by the adversary is modeled similarly to the rogue public key attack described in Section~\ref{AttackModels:RogueAttack}.

\section{Evaluation}\label{Sec:Evaluation}

In this section, we 
provide guidance on model selection and identify
which models should be used for
which modelling scenarios. 
We have also carefully checked that the verification
of different aggregate signatures behaves as expected. We do not
report on that in detail, but rather refer the reader
to~\cite{hofmeierFormalizingAggregateSignatures2021} and the experiments at \cite{hofmeierTamarinModelsAggregate}.

\looseness=-1
We first examine the differences between 
the attack-finding and validation models, as well as between the different versions of these models, by comparing them on a simple protocol.
We also evaluate the proof times and attack-finding times
on that protocol, illustrating the termination issues we
encountered. 
Together, this provides the modeler with some intuition on which models should be used for their verification efforts.

\subsection{Comparing the Models}\label{Sec:Evaluation:Comparing}

The validation models allow for arbitrary undesirable behaviors by providing the adversary with non-honest keys.
In contrast, the standard attack-finding model does not allow for any undesirable behaviors and thus models a primitive with much stronger properties.
Thus, proofs with the validation models provide stronger guarantees than proofs with the attack-finding models.
However, there are also reasons to use the attack-finding models.
First, protocols might rely on primitives with stronger guarantees than the primitives modeled by the validation models.
Second, as we will see in Section~\ref{Sec:Eval:Experiments}, achieving termination with the verification models is more difficult than with the attack-finding models.
Thus, weaker proof guarantees can be chosen over non-ter\-mi\-na\-tion. 
And finally, as the name \emph{attack-finding model} suggests, these models can be used to identify the behavior of the signature that allows for the attack.

Our extensions for the rogue public key attack are intended to 
model BLS aggregate signatures that mitigate the rogue public key attack on the protocol layer, such as Proof of Possession.
The other mitigation techniques mitigate the attack as part of the primitive's design
and thus the mitigation can be assumed to be effective.

As mentioned, we highlight the differences of our attack-finding models and validation models with a simple protocol.
Signers sign fresh messages with their secret keys and send these messages and signatures to an aggregator who aggregates an arbitrary number of these signatures and sends the signature aggregation to the verifier who looks up the signers' keys in the \ac{PKI} and verifies the aggregate signature.
We compare the model's properties using four lemmas:

\begin{LaTeXdescription}
  \item[Message authenticity] For each message $m$, signed with a key of an honest signer $S$ and received by the honest verifier $V$, $S$ must have signed $m$.
  \item[Weak agreement] As defined by Lowe~\cite{loweHierarchyAuthenticationSpecifications1997}, this property holds when the verifier $V$ verifies a signature for a signer $S$, then $S$ intended the message for $V$. This property does not hold trivially, as the signer does not include any information on the verifier in the signature. 
  \item[No splitting zero attack] An aggregate signature $\sigma_\text{agg}$ cannot be validated twice using different messages $\mathbf{m}_a$ and $\mathbf{m}_b$.
  \item[No rogue key attack] An honest verifier cannot verify an aggregate signature for a message $m$ and the public key of an honest signer $S$, when $S$ did not sign $m$.
\end{LaTeXdescription}

We summarize the results in Table~\ref{table:AttackFinding:Results} and refer to the Models~1-6 based on the numbers assigned in this table.
We compare the attack-finding Model~4 without additional adversary capabilities, as well as the two attack-finding Models~5 and 6 that support colliding signatures and rogue public key attacks with three versions of the validation model: the validation Models~1 and 2 with and without non-honest keys and the validation Model~3 supporting the rogue public key attack.
Note that these results come from simplified models due to termination issues, see Section~\ref{Sec:Eval:Experiments} for details.

The validation Model~3 with rogue public key attacks provides the weakest properties with respect to our lemmas.
And the attack-finding Model~4 and validation Model~1 without non-honest keys provides the strongest properties with respect to our lemmas. 
As mentioned above, the undesirable behaviors of the validation models are allowed by providing the adversary with non-honest keys.
Thus, the validation Model~1 without non-honest keys does not allow for colliding signatures and thus does not violate the \emph{no splitting zero attack} lemma, whereas the validation Model~2 with non-honest keys allows for this undesirable behavior and violates the lemma.

The attack-finding Model~5 with colliding signatures and the validation Model~2 with non-honest keys have the same verification results with respect to our lemmas. 
What the table does not show is that the validation Model~2 covers various attacks without stating them explicitly.
For example, there is an attack on the \ac{DEO} property, known from~\cite{porninDigitalSignaturesNot2005}, which is possible with the validation Model~2 but not with the attack-finding Model~5.
In other words, we can formulate additional properties that are falsified for Model~2 and are proven for Model~5.

\setlength{\arrayrulewidth}{0.25mm}
\begin{table}[]
\caption{Proven and falsified lemmas for all models}
  \centering
	\begin{tabular}{p{0.08cm}p{2.13cm}|p{0.9cm}|p{0.9cm}|p{1cm}|p{1.07cm}|}
		\hhline{~~|----}
		&  & \multicolumn{4}{c|}{Lemma} \\ \hline
    \multicolumn{2}{|c|}{Model} & Message authenticity& Weak Agreement &No splitting zero attack & No rogue key attack \\ \hline
    \multicolumn{2}{|p{2.658cm}|}{\cellcolor[HTML]{91bbed}\textbf{Validation models} }& &  & &  \\ \hline
    \multicolumn{1}{|p{0.08cm}|}{1} & \cellcolor[HTML]{91bbed}No non-honest keys& \cellcolor[HTML]{4ce056}Proven & \cellcolor[HTML]{fa837f}Falsified & \cellcolor[HTML]{4ce056}Proven & \cellcolor[HTML]{4ce056}Proven \\ \hline
    \multicolumn{1}{|p{0.08cm}|}{2} & \cellcolor[HTML]{91bbed}With non-honest keys & \cellcolor[HTML]{4ce056}Proven & \cellcolor[HTML]{fa837f}Falsified & \cellcolor[HTML]{fa837f}Falsified  & \cellcolor[HTML]{4ce056}Proven  \\ \hline
    \multicolumn{1}{|p{0.08cm}|}{3} & \cellcolor[HTML]{91bbed}Rogue public key & \cellcolor[HTML]{fa837f}Falsified & \cellcolor[HTML]{fa837f}Falsified & \cellcolor[HTML]{fa837f}Falsified & \cellcolor[HTML]{fa837f}Falsified \\ \hline
		\multicolumn{2}{|p{2.658cm}|}{\cellcolor[HTML]{dfb6e3}\textbf{Attack-finding models}} & &  & &  \\ \hline
		\multicolumn{1}{|p{0.08cm}|}{4} & \cellcolor[HTML]{dfb6e3}No adversary capabilities & \cellcolor[HTML]{4ce056}Proven & \cellcolor[HTML]{fa837f}Falsified & \cellcolor[HTML]{4ce056}{Proven}& \cellcolor[HTML]{4ce056}{Proven} \\ \hline
		\multicolumn{1}{|p{0.08cm}|}{5} & \cellcolor[HTML]{dfb6e3}Colliding signatures & \cellcolor[HTML]{4ce056}Proven & \cellcolor[HTML]{fa837f}Falsified & \cellcolor[HTML]{fa837f}Falsified & \cellcolor[HTML]{4ce056}Proven \\ \hline
		\multicolumn{1}{|p{0.08cm}|}{6} & \cellcolor[HTML]{dfb6e3}Rogue public key & \cellcolor[HTML]{fa837f}Falsified & \cellcolor[HTML]{fa837f}Falsified & \cellcolor[HTML]{4ce056}Proven & \cellcolor[HTML]{fa837f}Falsified \\ \hline
	\end{tabular}
  	\label{table:AttackFinding:Results}
\end{table}

\subsection{Experiments}\label{Sec:Eval:Experiments}
\begin{table*}[]
	\centering
	\caption{Proof and trace finding times in seconds for the simplified and unsimplified models. The measurements represent the CPU time of a single proof or falsification. Measurements made with Tamarin's development version compiled at 05.03.2021 on an Intel Xeon 2.20GHz 48 core computer with 256GiB RAM. The time out is set to one hour.}
	\begin{tabular}{p{0.25cm}p{4cm}|p{2.7cm}|p{1.8cm}|p{1.6cm}|p{1.9cm}|p{1.6cm}|}
		\hhline{~~|-----}
		& &Executable two signatures (exists-trace) & Message auth-enticity (prove) & Weak agreement (falsify)& No splitting zero attack (prove)& No rogue key attack (prove)\\ \hline
    \multicolumn{1}{|l}{1}&\multicolumn{1}{|l|}{\cellcolor[HTML]{91bbed}\textbf{Validation} \textbf{model}}& & & & & \\ \hline
		\multicolumn{1}{|l}{1a}&\multicolumn{1}{|l|}{\cellcolor[HTML]{91bbed}No simplifications} & \cellcolor[HTML]{EDE666}13.8 s & \cellcolor[HTML]{F37B70}Time out & \cellcolor[HTML]{EDDA66}23.3 s& \cellcolor[HTML]{F37B70}Time out & \cellcolor[HTML]{F37B70}Time out \\ \hline
		\multicolumn{1}{|l}{1b}&\multicolumn{1}{|l|}{\cellcolor[HTML]{91bbed}Limit number of signatures to three} & \cellcolor[HTML]{EDE666}14.3 s & \cellcolor[HTML]{EDCE66}1.2 min& \cellcolor[HTML]{EDE666}19.6 s& \cellcolor[HTML]{EDB066}4.9 min& \cellcolor[HTML]{EDCE66}1.0 min \\ \hline
		\multicolumn{1}{|l}{4}&\multicolumn{1}{|l|}{\cellcolor[HTML]{dfb6e3}\textbf{Attack-finding} \textbf{model}} & & & & & \\ \hline
		\multicolumn{1}{|l}{4a}&\multicolumn{1}{|l|}{\cellcolor[HTML]{dfb6e3}No simplifications} & \cellcolor[HTML]{BEDA6D}1.2 s & \cellcolor[HTML]{F37B70}Time out & \cellcolor[HTML]{EDEC66}8.5 s & \cellcolor[HTML]{88DA6D}0.2 s & \cellcolor[HTML]{F37B70}Time out \\ \hline
		\multicolumn{1}{|l}{4b}&\multicolumn{1}{|l|}{\cellcolor[HTML]{dfb6e3}Limit number of signatures to three} & \cellcolor[HTML]{BEDA6D}1.2 s& \cellcolor[HTML]{EDE666}10.3 s& \cellcolor[HTML]{D2DC5F}5.2 s & \cellcolor[HTML]{88DA6D}0.2 s & \cellcolor[HTML]{A1DA6D}1.1 s \\ \hline
		
	\end{tabular}
	\label{table:evaluation:compare simplifications}
\end{table*}

In this section, we evaluate how well our models can be used for verification and falsification in practice, with reasonable analysis times.
We compare the basic validation Model~1
and the basic attack-finding Model~4, both without additional adversary capabilities as
they have the same verification results with respect to our lemmas;
this allows us to compare their verification and falsification times with respect to each lemma.
We evaluate the models on the above simple protocol and lemmas.
We present the results in Table~\ref{table:evaluation:compare simplifications}.

Our original models,
presented in the Rows 1a and 4a,
did not terminate for the proof search for the \emph{message authenticity} and \emph{no rogue key attack} lemmas.
The non-termination probably stems from the repeated application of rules.
These rules enable arbitrarily many aggregated signatures
by modeling the aggregation and key lookup using loops, 
similar to the aggregation rules in Section~\ref{Sec:AttackFindingModels}. 

We apply and evaluate different methods to improve proof times.
Here we discuss the most effective method, which is limiting the number of aggregated signatures with restrictions.
For more details see~\cite{hofmeierFormalizingAggregateSignatures2021}.
Table~\ref{table:evaluation:compare simplifications} shows the results of evaluating these simplifications on the validation Model~1b,
and the attack-finding Model~4b.
Limiting the number of aggregated signatures directly limits Tamarin's search space as aggregating signatures beyond the stated limit is not considered.
Thus, it is not surprising that with this method the proofs of all the previously non-terminating lemmas terminate.
We also presented these results in Section~\ref{Sec:Evaluation:Comparing}.
The drawback of this technique is that we can only prove properties for this limited number of signatures and we may miss attacks when more signatures are aggregated than was analyzed.
However, if an attack is possible for a small number of signatures, this simplification can be very effective for finding attacks.

The fundamental difference between the attack-finding models and validation models is that the attack-finding models explicitly model the allowed verifications while the validation models disallow unwanted behaviors.
This should make the search space for the validation models larger than the search space for the attack-finding models.
This is confirmed by our experiments.
Note that Tamarin can prove the \emph{no splitting zero attack} lemma in under 0.2 seconds for the attack-finding Model~4a since this attack directly contradicts the aggregation equation.

\section{Case Study: SANA}\label{Sec:SANA}

In the previous section, we evaluated our aggregate signature models on a simple, artificial protocol.
In this section, we demonstrate that our models can be used to analyze substantially larger and more complex protocols.

We analyze the Scalable Aggregate Network Attestation (SANA) protocol \cite{ambrosinSANASecureScalable2016}. 
SANA attests whether devices' software state is
good, i.e. it matches the latest non-compromised software version, particularly when many devices are 
on the same network
and are expected to be in identical states.
This is helpful in contexts like 
building automation and IoT. 
There, devices could be responsible for access control, e.g.,
limiting the access of unauthorized entities to 
the building.
If the attestation protocol is insecure,
compromised devices 
may remain undetected and unidentified. This would
allow an attacker to enter restricted areas of the building, thereby violating the buildings' security policy. 

As SANA's authors omit some critical details, we analyze SANA under a range of reasonable assumptions.
Tamarin finds attacks on SANA for some of these assumptions while proving the protocol to be secure under others,
highlighting the importance of these details.
We disclosed the issues we found to SANA's authors.
All our models are available at~\cite{hofmeierTamarinModelsAggregate} and additional details can be found in the bachelor thesis~\cite{ragusoFormalAnalysisNetwork2023}.

\subsection{The SANA Protocol}
\label{sec:SANA_description}

One central concern of SANA is separating the act of attesting the software state of devices from other administrative duties.
The software attestation is carried out by one of potentially many verifying parties.
In contrast, the network is administrated by the network owner. 
To accommodate this separation, SANA is divided into two subprotocols.
The \tokenrequest{} subprotocol is an offline authorization protocol, 
where a verifier obtains a signed authorization token from the network owner.
Afterwards, this verifier can use said token to run the \attestation{} subprotocol,
which checks whether the devices on the network are running approved software.

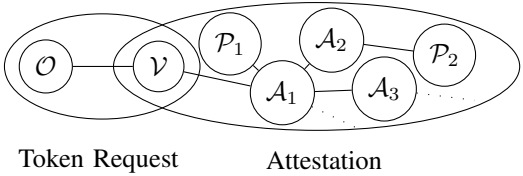
\begin{figure}
    \centering
	\tikzset{int/.style  = {draw, circle, minimum size=0.1em}}
		\begin{tikzpicture}[auto,>=latex]
	  \node at (-0.055, 0.1) [int] (a) {$\mathcal A_1$};
	  \node at (-1.7, 0.45) [int] (v) {$\V$};
	  \node at (-3.2, 0.45) [int] (o) {$\Ow$};
	  \path[-] (o) edge node[below] {} (v);
	  \path[-] (v) edge node[above] {} (a);
	  \node at (0.6, 0.8) [int] (a2) {$\mathcal A_2$};
	  \node at (1.3, 0.15) [int] (a3) {$\mathcal A_3$};
	  \node at (-0.75, 0.75) [int] (p1) {$\mathcal P_1$};
	  \path[-] (a) edge node[above] {} (a2);
	  \path[-] (a) edge node[above] {} (a3);
	  \path[-] (a) edge node[above] {} (p1);
	  \node at (2.1, 0.6) [int] (p2) {$\mathcal P_2$};
	  \path[-] (a2) edge node[above] {} (p2);
	  \node at (2.5, 0) (c) [coordinate] {a};
	  \node at (1, -0.4) (d) [coordinate] {a};
	  \path[loosely dotted] (a3) edge node[above] {} (c);
	  \path[loosely dotted] (a) edge node[above] {} (d);
	  \draw (-2.45,0.45) ellipse (1.3cm and 0.7cm);
	  \draw (0.4, 0.45) ellipse (2.7cm and 0.85cm);
	  \node at (-2.5, -0.8) {Token Request};
	  \node at (0.5, -0.8) {Attestation};
	\end{tikzpicture}
		\caption{Overview of SANA's aggregation tree}
	\label{fig:SANA:Overview}
\end{figure}

More specifically, the devices participating in SANA have one or more of the roles described below.
Figure~\ref{fig:SANA:Overview} provides an overview of the role assignment of one protocol run.
\begin{LaTeXdescription}
  \item[Verifier \V] initiates both the \tokenrequest{} and \attestation{} subprotocols and 
  verifies the attestation result. A verifier possesses an asymmetric key pair $(\skV, \pkV)$. 
  Multiple verifiers may be present in the network, 
  but only one participates in any given protocol run.
  \item[Provers $\mathcal P_i$] are the devices subject to software attestation. 
  When challenged,
  they compute an attestation response that is a signed hash over their software.
  Provers execute all protocol code in a tamper-proof secure execution environment. 
  Thus, we assume that all provers behave honestly in our analysis.
  Each prover is initialized in a trusted environment, 
  where it obtains an asymmetric key pair $(sk_i, pk_i)$ 
  and an authentic copy of the network owner's public key $\pkO$.
  \item[Network Owner \Ow] initializes the provers.
  In \tokenrequest{}, the owner authorizes a verifier $\V$ to attest the network.
  It has knowledge of all devices present in the network and their public keys ($\pkV$ and all $pk_i$). 
  Owners are assumed to be honest in~\cite{ambrosinSANASecureScalable2016}.
  \item[Aggregators $\mathcal A_i$] distribute $\V$'s messages to provers and other aggregators 
  and aggregate the provers' attestation responses.
  This aggregation is presented to $\V$ as the attestation result.
\end{LaTeXdescription}

\subsubsection{Optimistic Aggregate Signatures}\label{sec:SANA:OAS}
While aggregate signatures reduce storage and bandwidth requirements, 
they usually do not reduce the verification time complexity. 
Thus, the authors of SANA propose \emph{Optimistic Aggregate Signatures} (OAS), 
a variant of traditional aggregate signatures.
As 
most provers are 
expected to sign the same message, OAS reduce the verification complexity by specifying 
a default message $M$ and a separate aggregation of the public key.

More specifically, an OAS scheme is a tuple of algorithms 
(\oaskgen{}, \oasaggpk{}, \oassgn{}, \oasaggsgn{}, \oasvfy{}),
where \oaskgen{}, \oassgn{}, and \oasaggsgn{} are defined as for traditional aggregate signatures.
\oasaggpk{} aggregates multiple public keys into an aggregate public key.
\oasvfy{} provides more information compared to its traditional counterpart.
Given an aggregate signature, an aggregate public key, and a default message $M$,
\oasvfy{} either outputs $\bot$ upon failure or otherwise 
a set $\mathcal B$ that for each signed message $m_i \neq M$ contains a tuple $(m_i, S_i)$, where $S_i$ is the set of public keys whose corresponding secret keys were used to sign $m_i$.
\paper{The formal definition of an OAS scheme and its correctness and unforgeability notions are given in this paper's extended version~\cite{extendedVersion}.}\extended{The formal definition of an OAS scheme and its correctness and unforgeability notions are given in Appendix~\ref{OAS_corr_unforg}.}

No formal models of OAS schemes exist at the time of writing.
However, the proposed OAS scheme provides a similar functionality 
to traditional aggregate signatures, 
whereby the main difference lies in an optimized verification algorithm.
Since the performance of cryptographic algorithms 
is usually not accounted for when analyzing protocols in the symbolic model, 
the described OAS scheme can be modeled 
with our traditional aggregate
signature models.

\subsubsection{The Token Request Subprotocol}

\tokenrequest{} is executed between a verifier and the network owner 
and authorizes the verifier to execute a single instance of the \attestation{} subprotocol. 
This allows for
the attestation to be a public service while mitigating DoS attacks.

A simplified version of \tokenrequest{} is depicted in Figure~\ref{fig:tokenReq}. 
The signatures exchanged in this subprotocol are generated with a normal digital signature scheme, not the OAS scheme introduced above.
To initiate a protocol run, a verifier $\V$ samples a random nonce $N_\V$ and sends it to the network owner $\Ow$, 
which responds with its own nonce $N_\Ow$.
Next, $\V$ computes and sends a signature over $N_\Ow$ together with a public key certificate \certComp$(\pkV)$, 
issued by a trusted third party, to $\Ow$.
$\Ow$ then assembles and signs the authorization token $T$, 
which consists of a set $\mathcal H$ of hash values of all valid software configurations, 
a counter value $c$, which provides replay protection, 
and an expiry time\-stamp $t$. $\Ow$ aggregates the provers' public keys $pk_1, \dots, pk_n$ into the aggregate public key $apk$ and 
sends the following back to $\V$: The token~$T$, encrypted under $\pkV$, $apk$, a signature over $apk$ and $N_\V$, 
and a public key certificate \certComp$(\pkO)$.
Finally, $\V$ decrypts the token.
Naturally, if the verification of any of the exchanged signatures fails, 
protocol execution is aborted.

\begin{figure}
  \centering
  \scalebox{0.9}{
  \begin{msc}[label distance=0.04cm, msc keyword=, instance distance = 4.2cm, environment distance = 1.0cm, right environment distance=0cm, left environment distance=0cm, last level height=0.2cm, foot height=0.1cm, foot distance=0cm, head height=0.4cm]{}
    \setlength{\instwidth}{1.5cm}
    \setlength{\topheaddist}{0.2cm}
    \setlength{\firstlevelheight}{0.2cm}
    \drawframe{no}
    \declinst[label distance=0.1cm]{V}{$\skV, \pkO$}{$\V$}
    \declinst[label distance=0.1cm]{O}{$\skO, \pkV$}{$\Ow$}
    \mess{$N_\V \gets_\$ \{0,1\}^{l_N}$}{V}{O}
    \nextlevel[1]
    \mess{$N_\Ow \gets_\$ \{0,1\}^{l_N}$}{O}{V}
    \nextlevel[1]
    \mess{\parbox{3cm}{$\signComp(N_\Ow, \skV), \certComp(\pkV)$}}{V}{O}
    \nextlevel[2.8]
        \mess{\parbox{5.3cm}{$\encComp(\overset{T}{\overbrace{\left(\mathcal H, c, t, \signComp(\mathcal H | c |t, \skO)\right)}}, \pkV), $ \\ $apk, \signComp(N_\V | apk, \skO), \certComp(\pkO)$}}{O}{V} 
  \end{msc}
  }
  \caption{Simplified SANA \tokenrequest{} subprotocol}  
  \label{fig:tokenReq}
  \end{figure}

  \subsubsection{The Attestation Subprotocol}

  This subprotocol uses the result of the prior \tokenrequest{}
  as a starting point. It is intended to provide the verifier
  with guarantees that all provers are in a good state, i.e. their software version matches the latest non-compromised software version. 

  The \attestation{} subprotocol follows a simple challenge-response mechanism between the verifier and the provers.
  The verifier assembles an attestation challenge $Ch = \{N, T\}$, 
  consisting of the previously obtained authorization token~$T$ and a fresh nonce~$N$.
  To distribute $Ch$, the verifier sends it to a gateway aggregator, which forwards it recursively to its adjacent aggregators and provers.
  This induces a tree-shaped communication path over the network, called an \textit{aggregation tree}, see Figure~\ref{fig:SANA:Overview}.
  Before forwarding $Ch$, an aggregator validates it by verifying the signature and counter in $T$.
    The authors claim to mitigate DoS attacks with this mechanism.
  A prover, upon receiving a challenge, first validates the token contained therein
    and then computes a hash value $h$ over its software.
  It then signs the message $m := h | N | c$ using the OAS scheme and returns the signature to its parent in the aggregation tree.
  Aggregators collect and aggregate all received signatures and send them back towards the root of the aggregation tree. 
  The gateway aggregator presents the aggregation of all responses to the verifier,
  which then verifies it against the aggregate public key received from the network owner during \tokenrequest.
  Note that the generation of hashes for software attestation is out of SANA's scope.

\subsection{Security Properties}
We first present the security properties we considered when analyzing the \tokenrequest{} subprotocol in isolation 
and then present the security properties for the combination of both subprotocols. 

\subsubsection{Token Request Properties}
  The authors of SANA~\cite{ambrosinSANASecureScalable2016} do not state any concrete security properties to be satisfied by \tokenrequest.
  They only state properties for the whole attestation scheme, given below.
  Therefore, we make minimal assumptions and 
  evaluate what properties the protocol satisfies.
  Specifically, since \tokenrequest{} serves as an authorization mechanism, we analyze the protocol with respect to the increasingly strict notions of entity authentication, introduced by Lowe~\cite{loweHierarchyAuthenticationSpecifications1997}.

\subsubsection{Attestation Properties}
We consider two properties.

\begin{LaTeXdescription}
  \item[Attestation Agreement.]
  This is a formalization of what SANA's authors call \emph{Unforgeability and Freshness}, which they define as follows:
  If an honest verifier was able to verify an attestation result including a given uncompromised prover, then the claimed integrity measurement reflects the state of the prover's software during the protocol run. 
  We formalize this property as non-injective agreement between the verifier and any prover, parameterized on the prover's attestation response.
          \item[Token Agreement.]
    This states that a prover should act upon receiving a token only if that token was issued by the owner.
  We formalize this as non-injective agreement between the owner and a given prover, parameterized on the token.
\end{LaTeXdescription}

While the first property is stated as part of the main requirements for SANA in~\cite{ambrosinSANASecureScalable2016}, 
we include the second in our analysis to validate the effectiveness of \tokenrequest{} as an authorization mechanism.

\subsection{Modelling Assumptions}
\label{sec:SANA:model_assumptions}

\looseness=-1
  The authors of \cite{ambrosinSANASecureScalable2016} make the assumption
  that each device in the network can identify and communicate with its direct neighbors.
  Moreover, the network's topology 
  may change dynamically.
    To simplify our analysis and to account for topology changes, we assume a fully connected network, where each device can communicate with every other network device. 
  We also assume the adversary has full control over the network, allowing the adversary, for example, to delay or drop messages.
  This is standard in the symbolic model, and ensures that verification results are valid in practice, even in the worst case. 

  Furthermore, the authors of SANA mandate an existing trusted third-party public key infrastructure (PKI) used by $\Ow$ and $\V$,
  which authenticates the public keys $\pkO$ and $\pkV$ using signed certificates.
  As the PKI's capabilities are not further specified, we employ 
  a standard model of a PKI, where the PKI and certificates are abstracted as a public dictionary that binds public keys to identities.
  As this PKI is independent of SANA, we assume it is unaware of SANA's roles and thus, it does not bind the parties' protocol roles to their public key.

  As stated in Section~\ref{sec:SANA_description}, 
  each prover is initialized with an authentic copy of the owner's public key.
        In contrast, the initialization of the \emph{verifiers} is not described by SANA's authors.
  Thus, we analyzed the protocol under two different assumptions:
  First, assuming that verifiers are only initialized with their own secret key $\skV$ and, 
  second, additionally assuming the initialization with the owner's identity to look up its public key in the PKI.
  The first assumption is less restrictive and is based on
  SANA being motivated as a public attestation service allowing anyone to request an authorization token.
    Thus, we first assume that verifiers have no prior initialization with the owner.
  Consultation with the SANA authors, however, revealed the implicit assumption that the owner's identity is known.
  This motivates our second assumption about verifiers being initialized with the owner's identity.
  
  The standard PKI model allows the adversary to compromise individual secret keys.
  Note that an agent loses most security guarantees when its secret key is compromised.
  Thus, security properties only consider the guarantees provided to non-compromised, i.e., honest, agents that act in the presence of compromised agents.
  The authors of SANA also assume the owner to be honest.
  This assumption is indeed needed, as otherwise Tamarin finds trivial attacks where the adversary uses the owner's secret key to issue a fake authorization token.

\subsection{Results}\label{sec:SANA:Results}
We first present the results of analyzing the \tokenrequest{} subprotocol separately and then the results of our analysis of the combination of both subprotocols.

\subsubsection{Token Request Analysis}
\label{sec:SANA:tokenrequest:results}
\begin{table}[]
	\caption{Results for the \tokenrequest{} subprotocol.}
  \centering
	\begin{tabular}{|l|l|l|l|}
    \hhline{~~|--}
    \multicolumn{1}{c}{} & \multicolumn{1}{c|}{} & \multicolumn{2}{c|}{Initialization $\V$}\\
    \hline
    Property & Subject & None & Owner Identity \\ \hline
    \multirow{2}{4.2em}{Aliveness} & Owner & \cellcolor[HTML]{4ce056}Proven & \cellcolor[HTML]{4ce056}Proven \\ \hhline{|~|---}
     & Verifier & \cellcolor[HTML]{fa837f}Falsified & \cellcolor[HTML]{4ce056}Proven \\ \hline
    \multirow{2}{4.2em}{Weak Agreement} & Owner & \cellcolor[HTML]{fa837f}Falsified & \cellcolor[HTML]{fa837f}Falsified \\ \hhline{|~|---}
     & Verifier & \cellcolor[HTML]{fa837f}Falsified & \cellcolor[HTML]{fa837f}Falsified \\ \hline
	\end{tabular}
	\label{table:tokenrequest:results}
\end{table}

The results of our analysis of \tokenrequest{} are summarized in Table~\ref{table:tokenrequest:results}, where the subject denotes the party to which the property is guaranteed.
We note that only the weakest two notions of entity authentication, Aliveness and Weak Agreement, need to be checked, as Weak Agreement fails to hold, as explained below, and thus all stronger properties also fail.

When assuming that the verifier is \emph{not} initialized with the owner's identity, 
Tamarin finds counterexamples for all notions of entity authentication except for Aliveness of the verifier from the perspective of the owner.
Assuming verifier initialization with the owner's identity the protocol
additionally provides Aliveness of the owner.
We provide the three counterexamples corresponding to the weakest notions of authenticity 
\paper{in this paper's extended version~\cite{extendedVersion}.}\extended{in Appendix~\ref{SANA:attestation:attacks}.}
Here, we highlight the flaws that enable these attacks.

The \tokenrequest{} subprotocol provides a signing oracle that allows an adversary to obtain a valid signature over an arbitrary term under any verifier's secret key, provided this verifier initiates an execution of \tokenrequest{}.
The adversary can provide any message instead of the nonce~$N_\Ow$, which the verifier will then sign and provide to the adversary.
This is possible as this term is not cryptographically tied to the rest of the protocol execution.
In the counterexample for Aliveness from the verifier's perspective, the adversary obtains the apparent owner's signatures over both the aggregate public key and the token using the signing oracle.
Thus, the targeted verifier $\V$ obtains a forged token without the owner's involvement.
Note that this attack relies on the fact that $\V$ accepts another verifier's public key $pk_{\V'}$ as an owner's public key.
Thus, the initialization of the verifier with the owner's identity prevents this attack.

\looseness=-1
Tamarin also finds attacks on Weak Agreement from the perspective of both the owner and verifier, for both initialization assumptions.
These attacks exploit the fact that the exchanged signatures do not authenticate the designated receiver.
Most notably, SANA's authorization token, 
which authorizes
a specific verifier to perform network attestation, 
is not bound to that verifier.
In the attack from the perspective of the verifier, 
the targeted verifier receives a token that was intended for another (e.g. the adversary).
This attack shows that the adversary can forward 
a token to an arbitrary verifier, even if the owner would 
ordinarily refuse to 
issue
a token to the latter.

The counterexample for the Weak Agreement property, from the perspective of the network owner, exploits that the verifier's signature over the owner's nonce does not bind the owner identity to this protocol run.
The attack results in the owner and verifier disagreeing on their communication partner.
Namely, the adversary lets the owner believe that a verifier is communicating with them, while the verifier believes it is communicating with a different owner.
This would require that verifiers may attest networks with different owners.
It is unclear whether SANA would be deployed in such a scenario and thus whether such an attack could be exploited in practice.

Although the exact exploitation of these attacks is unclear, they clearly show that SANA's \tokenrequest{} only provides weak authentication properties.
Even with the stronger initialization assumptions, the protocol only guarantees Aliveness, i.e.,
the parties know that their communication partner executed the protocol at some point, 
but cannot be sure that their communication partner intends to communicate with them in any given protocol run.
This rather weak property is achieved by a four message protocol that involves fresh nonces.
Note, however, that a simpler two message protocol could achieve the same level of authentication and 
that slightly modifying the signed messages in the original protocol results in the much stronger Injective Agreement property, as described in~\cite{ragusoFormalAnalysisNetwork2023}.

\subsubsection{Complete Analysis}
\label{sec:SANA:attestation_results}

We present the results of our analysis of the combination of both subprotocols in Table~\ref{table:Attestation:Results}.
We consider four model classes. 
Each class consists of different models, each using a different aggregate signature model.
Table~\ref{table:Attestation:Results} specifies the used aggregate signature models, referring to the numbers from Table~\ref{table:AttackFinding:Results}.

The first two model classes assume standard adversary capabilities, as described in Section~\ref{sec:SANA:model_assumptions}.
As in Section~\ref{sec:SANA:tokenrequest:results}, we make two assumptions on the verifier initialization.
First, we assume the verifier is only initialized with its own key pair and, second, we assume the verifier is additionally initialized with the owner's identity.
The weakest of our aggregate signature models, Model~4, already 
results in an attack when assuming no verifier initialization.
The stronger assumption of initializing the verifier with the owner identity is verified by Tamarin for each of our six aggregate signature models, see Table~\ref{table:AttackFinding:Results}.
The second two model classes assume an additional adversary capability, where the adversary can register their non-honest keys with the owner to add them to the aggregate public key $apk$.
While this stronger adversary model is not very realistic for SANA, it showcases the impact of the rogue public key attacks.
Namely, Tamarin finds attacks for the two aggregate signature models with the rogue key attack, 
while Tamarin proves the properties for the other aggregate signature models.
Next, we present the results of these four model classes in more detail.

\begin{table}[]
  \caption{Results for the complete protocol analysis}
  \centering
  \begin{tabular}{p{0.9cm}p{1.25cm}|p{0.9cm}|p{1cm}|p{1.05cm}|p{0.9cm}|}
  \hhline{~~|----}
  & & \multicolumn{2}{p{2.3cm}|}{No Non-Honest Keys in $apk$} & \multicolumn{2}{p{2.3cm}|}{Non-Honest Keys in $apk$} \\ \hhline{~~|----}
  & & Init $\V$: None & Init $\V$: Owner Identity & No Rogue Keys & Rogue Keys \\ \hline
  \multicolumn{2}{|p{2cm}|}{Used Aggregate Signature Models} & Model 4 & Models 1-6 & Models 1, 2, 4, 5 & Models 3, 6 \\ \hline
  \multicolumn{1}{|p{0.9cm}|}{\multirow{1}{*}{Property}} & Attestation Agreement & \cellcolor[HTML]{fa837f}Falsified & \cellcolor[HTML]{4ce056}Proven & \cellcolor[HTML]{4ce056}Proven & \cellcolor[HTML]{fa837f}Falsified \\ \hhline{|~|-----}
  \multicolumn{1}{|p{0.9cm}|}{} & Token Agreement & \cellcolor[HTML]{4ce056}Proven & \cellcolor[HTML]{4ce056}Proven & \cellcolor[HTML]{4ce056}Proven & \cellcolor[HTML]{4ce056}Proven \\ \hline
  \end{tabular}
  	\label{table:Attestation:Results}
\end{table}

Tamarin proves the Token Agreement property for each of our four model classes.
This property is achieved through the authentication of the token by the owner and the fact that every prover is initialized with an authentic copy of the owner's public key.
Note that this \emph{prover} initialization is independent of our assumptions on the \emph{verifier} initialization.

For the model where the verifier is not initialized with the owner's identity, 
Tamarin finds counterexamples for the Attestation Agreement property.
This attack exploits the signing oracle in the \tokenrequest{} subprotocol described in Section~\ref{sec:SANA:tokenrequest:results} to obtain signatures for both a fake authorization token and one or more fake attestation responses.
This attack enables the adversary to present an arbitrary attestation result to the targeted verifier, i.e., it can make the targeted verifier believe that there are no devices with invalid software in the network while the attestation on the potentially compromised provers was never carried out.
\paper{See the extended version of this paper~\cite{extendedVersion} for more details on this attack.}\extended{See Appendix~\ref{SANA:attestation:attacks} for more details on this attack.}

\looseness=-1
The above attack relies on the adversary putting a verifier's public key into the aggregate public key of the forged token.
Thus, preventing the adversary from forging authorization tokens also prevents the above attack. 
This is achieved by assuming the verifier is initialized with the owner's identity, see Section~\ref{sec:SANA:tokenrequest:results}.
Accordingly, Tamarin proves Attestation Agreement for the model 
that take this assumption into account. 

Under this additional assumption, 
Tamarin proves Attestation Agreement for all six aggregate signature models.
At first glance it might appear surprising that the insecure and undesirable behaviors of aggregate signatures do not contribute to additional attacks against SANA.
However, as these behaviors rely on the presence of non-honest keys, which would need to be added to the $apk$,
they are prevented by the authentication of the aggregate public key $apk$.
As the provers are assumed to be initialized by the owner in a secure environment, it is reasonable to assume that the provers' keys are honest.
However, to showcase our aggregate signature models, we created models where the adversary can add non-honest keys to the aggregate public key $apk$.

With this additional adversarial capability, the aggregate signature models behave differently. 
Namely, the two examined properties are proven for the models without rogue public key attacks, Models~1, 2, 4, and 5.
Furthermore, Tamarin finds an attack on Attestation Agreement for the models \emph{with} the rogue public key attack, Models~3 and 6.
As expected, the adversary can forge an aggregate signature for a targeted prover and a corresponding rogue public key,
thus, again forging the attestation result.
Note that SANA is specified to have measures against the rogue public key attack implemented.
Thus, while showcasing our aggregate signature models, this attack is not realistic.
Furthermore, colliding signature attacks and similar undesirable behaviors do not enable attacks against the Attestation Agreement property 
as the latter requires the considered prover to be honest and have an honest key pair, which is a standard assumption.

This case study of the SANA protocol demonstrates how our different aggregate signature models can be used in practice.
It can be sufficient to use Model~4, which supports the fewest behaviors, to find attacks.
Models~1-6 can be used to determine necessary assumptions or to prove the security of the protocol.
Our case study also highlights the importance of the assumptions on keys: 
As the insecure and undesirable behaviors of aggregate signatures are based on non-honest keys, some assumptions on keys allow for attacks, while others prevent them.

\subsection{Model Limitations}
\label{sec:SANA:limitations}

First, as with all formal methods, the translation of a protocol specification from natural language to a formal model relies on our interpretation of the (informal) specification.
We confirmed some of our assumptions with SANA's authors.
Our second assumption on the verifier initialization, described in Section~\ref{sec:SANA:model_assumptions}, is based on this exchange.

\looseness=-1
Second, various parts of the SANA protocol have an iterative and potentially unbounded characteristic, e.g., the aggregation of attestation responses.
This resulted in the verification process aborting due to resource exhaustion in Tamarin.
As a result, we considered only a small number of provers in our models, namely, two or three (depending on the model).
Additional vulnerabilities may arise when considering more or even an unbounded number of provers.
The presented attacks for small numbers of provers of course extend to larger numbers.
However, the security claims are limited by such bounds.

\looseness=-1
Finally, we performed several protocol simplifications.
For a detailed discussion of all simplifications with arguments for their soundness, see \cite{ragusoFormalAnalysisNetwork2023}.
While our attacks should be applicable in an actual deployment, we could not confirm this, as no implementation of the SANA protocol is publicly available.

\section{Conclusion}\label{Sec:Conclusion}

We developed the first symbolic models of aggregate signatures for Tamarin.
We created two classes of models: validation models based on the computational definition of aggregate signatures, and attack-finding models based on Tamarin's built-in signatures.
These classes of models are inspired by the signature models by Jackson et~al.~\cite{jacksonSeemsLegitAutomated2019}.

Our generic models are based on the general definition of aggregate signatures, and to model BLS aggregate signatures we extended our general aggregate signature models with known attacks on BLS.
We extended the attack-finding model for colliding signatures on aggregate signatures and we extended both classes of models for the rogue public key attack.

Our new aggregate signature models enable the analysis of protocols that rely on aggregate signatures,
which we demonstrate with the SANA protocol~\cite{ambrosinSANASecureScalable2016}.
Our analysis revealed important conditions that must be met during agent initialization to prevent attacks.
These conditions were not explicitly stated by the authors of SANA.

The arbitrary number of aggregated signatures, and the loops required to model such aggregations, make Tamarin's termination challenging.
Thus, some model simplifications were required for the proofs of some properties to terminate.
When using our models in larger protocols, proof techniques, such as oracles and inductive reuse lemmas, tailored to the protocol in question could help with termination.
Other options to improve termination could involve dedicated support in Tamarin, for example through improved heuristics.

Our techniques to model arbitrarily many aggregated signatures include the quantification over the elements in a multiset in lemmas and restrictions, using indices in multisets to represent vectors, and using dedicated adversary rules to transform signatures into an aggregation.
These techniques are general and could be applied to other multi-party primitives, such as threshold signatures or group signatures.

\bibliographystyle{IEEEtran}
\bibliography{aggSignatures}

\begin{thebibliography}{10}
\providecommand{\url}[1]{#1}
\csname url@samestyle\endcsname
\providecommand{\newblock}{\relax}
\providecommand{\bibinfo}[2]{#2}
\providecommand{\BIBentrySTDinterwordspacing}{\spaceskip=0pt\relax}
\providecommand{\BIBentryALTinterwordstretchfactor}{4}
\providecommand{\BIBentryALTinterwordspacing}{\spaceskip=\fontdimen2\font plus
\BIBentryALTinterwordstretchfactor\fontdimen3\font minus \fontdimen4\font\relax}
\providecommand{\BIBforeignlanguage}[2]{{%
\expandafter\ifx\csname l@#1\endcsname\relax
\typeout{** WARNING: IEEEtran.bst: No hyphenation pattern has been}%
\typeout{** loaded for the language `#1'. Using the pattern for}%
\typeout{** the default language instead.}%
\else
\language=\csname l@#1\endcsname
\fi
#2}}
\providecommand{\BIBdecl}{\relax}
\BIBdecl

\bibitem{eth2bookUpgradingEthereum}
B.~Edgington, ``{U}pgrading {E}thereum | 2.9.1 {B}{L}{S} {S}ignatures --- eth2book.info,'' \url{https://eth2book.info/capella/part2/building_blocks/signatures/}, 2023, [Accessed 27-04-2024].

\bibitem{drandCryptography}
Cloudflare, ``{C}ryptography --- drand.love,'' \url{https://drand.love/docs/cryptography/}, [Accessed 27-04-2024].

\bibitem{internetcomputerChainkeySignatures}
Dfinity, ``{C}hain-key signatures | {I}nternet {C}omputer --- internetcomputer.org,'' \url{https://internetcomputer.org/how-it-works/threshold-ecdsa-signing/}, [Accessed 27-04-2024].

\bibitem{bonehAggregateVerifiablyEncrypted2003a}
D.~Boneh, C.~Gentry, B.~Lynn, and H.~Shacham, ``Aggregate and {{Verifiably Encrypted Signatures}} from {{Bilinear Maps}},'' in \emph{Advances in {{Cryptology}} --- {{EUROCRYPT}} 2003}, E.~Biham, Ed.\hskip 1em plus 0.5em minus 0.4em\relax Berlin, Heidelberg: Springer, 2003, pp. 416--432.

\bibitem{boneh2001short}
D.~Boneh, B.~Lynn, and H.~Shacham, ``Short signatures from the weil pairing,'' in \emph{International conference on the theory and application of cryptology and information security}.\hskip 1em plus 0.5em minus 0.4em\relax Springer, 2001, pp. 514--532.

\bibitem{bonehDraftirtfcfrgblssignature05BLSSignatures2022}
\BIBentryALTinterwordspacing
D.~Boneh, S.~Gorbunov, R.~S. Wahby, H.~Wee, C.~A. Wood, and Z.~Zhang, ``Draft-irtf-cfrg-bls-signature-05 - {{BLS Signatures}},'' Jun. 2022. [Online]. Available: \url{https://datatracker.ietf.org/doc/draft-irtf-cfrg-bls-signature/05/}
\BIBentrySTDinterwordspacing

\bibitem{bonehGraduateCourseApplied}
\BIBentryALTinterwordspacing
D.~Boneh and V.~Shoup, \emph{A {{Graduate Course}} in {{Applied Cryptography}} - {{Version}} 0.5}, 2020. [Online]. Available: \url{http://toc.cryptobook.us/}
\BIBentrySTDinterwordspacing

\bibitem{lysyanskayaUniqueSignaturesVerifiable2002}
A.~Lysyanskaya, ``Unique {{Signatures}} and {{Verifiable Random Functions}} from the {{DH-DDH Separation}},'' in \emph{Advances in {{Cryptology}} --- {{CRYPTO}} 2002}, M.~Yung, Ed.\hskip 1em plus 0.5em minus 0.4em\relax Berlin, Heidelberg: Springer, 2002, pp. 597--612.

\bibitem{quan2021}
\BIBentryALTinterwordspacing
N.~T.~M. Quan, ``0,'' Tech. Rep. 323, 2021. [Online]. Available: \url{https://eprint.iacr.org/2021/323}
\BIBentrySTDinterwordspacing

\bibitem{blanchetProVerif03Automatic2021}
\BIBentryALTinterwordspacing
B.~Blanchet, B.~Smyth, V.~Cheval, and M.~Sylvestre, ``{{ProVerif}} 2.03: {{Automatic Cryptographic Protocol Verifier}}, {{User Manual}} and {{Tutorial}},'' Sep. 2021. [Online]. Available: \url{https://bblanche.gitlabpages.inria.fr/proverif/manual.pdf}
\BIBentrySTDinterwordspacing

\bibitem{schmidtAutomatedAnalysisDiffieHellman2012}
\BIBentryALTinterwordspacing
B.~Schmidt, S.~Meier, C.~Cremers, and D.~Basin, ``Automated {{Analysis}} of {{Diffie-Hellman Protocols}} and {{Advanced Security Properties}},'' in \emph{2012 {{IEEE}} 25th {{Computer Security Foundations Symposium}}}.\hskip 1em plus 0.5em minus 0.4em\relax Cambridge, MA, USA: IEEE, Jun. 2012, pp. 78--94. [Online]. Available: \url{http://ieeexplore.ieee.org/document/6266153/}
\BIBentrySTDinterwordspacing

\bibitem{ambrosinSANASecureScalable2016}
\BIBentryALTinterwordspacing
M.~Ambrosin, M.~Conti, A.~Ibrahim, G.~Neven, A.-R. Sadeghi, and M.~Schunter, ``Sana: Secure and scalable aggregate network attestation,'' in \emph{Proceedings of the 2016 ACM SIGSAC Conference on Computer and Communications Security}, ser. CCS '16.\hskip 1em plus 0.5em minus 0.4em\relax New York, NY, USA: Association for Computing Machinery, 2016, p. 731–742. [Online]. Available: \url{https://doi.org/10.1145/2976749.2978335}
\BIBentrySTDinterwordspacing

\bibitem{bonehDraftirtfcfrgblssignature04BLSSignatures}
\BIBentryALTinterwordspacing
D.~Boneh, S.~Gorbunov, R.~Wahby, H.~Wee, and Z.~Zhang, ``Draft-irtf-cfrg-bls-signature-04 - {{BLS Signatures}}.'' [Online]. Available: \url{https://datatracker.ietf.org/doc/draft-irtf-cfrg-bls-signature/04/}
\BIBentrySTDinterwordspacing

\bibitem{quanAttacksWeaknessesBLS2021}
\BIBentryALTinterwordspacing
N.~T.~M. Quan, ``Attacks and weaknesses of {{BLS}} aggregate signatures,'' Tech. Rep. 377, 2021. [Online]. Available: \url{https://eprint.iacr.org/2021/377}
\BIBentrySTDinterwordspacing

\bibitem{jacksonSeemsLegitAutomated2019}
\BIBentryALTinterwordspacing
D.~Jackson, C.~Cremers, K.~Cohn-Gordon, and R.~Sasse, ``Seems legit: Automated analysis of subtle attacks on protocols that use signatures,'' in \emph{Proceedings of the 2019 ACM SIGSAC Conference on Computer and Communications Security}, ser. CCS '19.\hskip 1em plus 0.5em minus 0.4em\relax New York, NY, USA: Association for Computing Machinery, 2019, p. 2165–2180. [Online]. Available: \url{https://doi.org/10.1145/3319535.3339813}
\BIBentrySTDinterwordspacing

\bibitem{BlanchetCSFW01}
B.~Blanchet, ``An {E}fficient {C}ryptographic {P}rotocol {V}erifier {B}ased on {P}rolog {R}ules,'' in \emph{14th IEEE Computer Security Foundations Workshop (CSFW-14)}.\hskip 1em plus 0.5em minus 0.4em\relax Cape Breton, Nova Scotia, Canada: IEEE Computer Society, Jun. 2001, pp. 82--96.

\bibitem{cryptoeprint:2024/1920}
\BIBentryALTinterwordspacing
C.~Cremers, A.~Peltonen, and M.~Zhao, ``An extended hierarchy of security notions for threshold signature schemes and automated analysis of protocols that use them,'' Cryptology {ePrint} Archive, Paper 2024/1920, 2024, [Accessed 15-04-2025]. [Online]. Available: \url{https://eprint.iacr.org/2024/1920}
\BIBentrySTDinterwordspacing

\bibitem{le-papinVerifyingListSwarm2023}
\BIBentryALTinterwordspacing
J.~{Le-Papin}, B.~Dongol, H.~Treharne, and S.~Wesemeyer, ``Verifying {{List Swarm Attestation Protocols}},'' in \emph{Proceedings of the 16th {{ACM Conference}} on {{Security}} and {{Privacy}} in {{Wireless}} and {{Mobile Networks}}}, ser. {{WiSec}} '23.\hskip 1em plus 0.5em minus 0.4em\relax New York, NY, USA: Association for Computing Machinery, Jun. 2023, pp. 163--174. [Online]. Available: \url{https://dl.acm.org/doi/10.1145/3558482.3581778}
\BIBentrySTDinterwordspacing

\bibitem{hofmeierFormalizingAggregateSignatures2021}
\BIBentryALTinterwordspacing
X.~Hofmeier, ``Formalizing {{Aggregate Signatures}} in the {{Symbolic Model}},'' Master's thesis, ETH Zurich, Department of Computer Science, Oct. 2021. [Online]. Available: \url{https://www.research-collection.ethz.ch/handle/20.500.11850/511820}
\BIBentrySTDinterwordspacing

\bibitem{meierTAMARINProverSymbolic2013}
S.~Meier, B.~Schmidt, C.~Cremers, and D.~Basin, ``The {{TAMARIN Prover}} for the {{Symbolic Analysis}} of {{Security Protocols}},'' in \emph{Computer {{Aided Verification}}}, ser. Lecture {{Notes}} in {{Computer Science}}, N.~Sharygina and H.~Veith, Eds.\hskip 1em plus 0.5em minus 0.4em\relax Berlin, Heidelberg: Springer, 2013, pp. 696--701.

\bibitem{tamarinProverManual}
Various, ``Tamarin prover manual,'' \url{https://tamarin-prover.com/manual/master/book/001_introduction.html}, [Accessed 27-04-2024].

\bibitem{hofmeierTamarinModelsAggregate}
\BIBentryALTinterwordspacing
X.~Hofmeier, A.~Raguso, R.~Sasse, D.~Jackson, and D.~Basin, ``One for all: Formally verifying protocols which use aggregate signatures,'' 2025, tamarin models of aggregate signatures and the SANA case study. [Online]. Available: \url{https://doi.org/10.5281/zenodo.15357177}
\BIBentrySTDinterwordspacing

\bibitem{loweHierarchyAuthenticationSpecifications1997}
\BIBentryALTinterwordspacing
G.~Lowe, ``A hierarchy of authentication specifications,'' in \emph{Proceedings 10th {{Computer Security Foundations Workshop}}}.\hskip 1em plus 0.5em minus 0.4em\relax Rockport, MA, USA: IEEE Comput. Soc. Press, 1997, pp. 31--43. [Online]. Available: \url{http://ieeexplore.ieee.org/document/596782/}
\BIBentrySTDinterwordspacing

\bibitem{porninDigitalSignaturesNot2005}
T.~Pornin and J.~P. Stern, ``Digital {{Signatures Do Not Guarantee Exclusive Ownership}},'' in \emph{Applied {{Cryptography}} and {{Network Security}}}.\hskip 1em plus 0.5em minus 0.4em\relax Berlin, Heidelberg: Springer, 2005, pp. 138--150.

\bibitem{ragusoFormalAnalysisNetwork2023}
\BIBentryALTinterwordspacing
A.~Raguso, ``Formal {{Analysis}} of {{Network Attestation}} and {{Aggregate Signatures}},'' Bachelor {{Thesis}}, ETH Zurich, Department of Computer Science, Mar. 2023. [Online]. Available: \url{https://www.research-collection.ethz.ch/handle/20.500.11850/731316}
\BIBentrySTDinterwordspacing

\bibitem{kapurEunificationAlgorithmAnalyzing2003}
D.~Kapur, P.~Narendran, and L.~Wang, ``An {{E-unification Algorithm}} for {{Analyzing Protocols That Use Modular Exponentiation}},'' in \emph{Rewriting {{Techniques}} and {{Applications}}}, R.~Nieuwenhuis, Ed.\hskip 1em plus 0.5em minus 0.4em\relax Berlin, Heidelberg: Springer, 2003, pp. 165--179.

\bibitem{bonehShortSignaturesWeil2001}
D.~Boneh, B.~Lynn, and H.~Shacham, ``Short {{Signatures}} from the {{Weil Pairing}},'' in \emph{Advances in {{Cryptology}} --- {{ASIACRYPT}} 2001}, ser. Lecture {{Notes}} in {{Computer Science}}, C.~Boyd, Ed.\hskip 1em plus 0.5em minus 0.4em\relax Berlin, Heidelberg: Springer, 2001, pp. 514--532.

\end{thebibliography}

\appendices

\extended{\section{Aggregate Signature Definitions}\label{Appendix:AggDefinitions}

In Section~\ref{Sec:Background:AggSign}, we described the definition of aggregate signatures, its attack game, and EUF-CMA security definition.
Here we provide these definitions in more detail according to Boneh and Shoup~\cite{bonehGraduateCourseApplied}.

We first provide the definition of an aggregate signature scheme according to Boneh and Shoup~\cite{bonehGraduateCourseApplied}.

\begin{definition}[Aggregate signature scheme~\cite{bonehGraduateCourseApplied}, page 623]\label{def:background:signatureAggregation}
	An aggregate signature scheme $\aggsignature = (\genComp, \signComp, \vfyComp, \aggComp, \vfyAggComp)$ is a signature scheme with two additional efficient algorithms $\aggComp$ and $\vfyAggComp$:
	\begin{itemize}
		\item A signature aggregation algorithm $\aggComp(\mathbf{pk}, \bm{\sigma})$ takes as input two equal length vectors, a vector of public keys $\mathbf{pk} = (pk_1, . . . , pk_n )$ and a vector of signatures $\bm{\sigma} = (\sigma_1, . . . , \sigma_n)$. It outputs an aggregate signature $\sigma_\text{agg}$.		
		\item The deterministic aggregate verification algorithm \newline $\vfyAggComp(\sigma_\text{agg}, \mathbf{m}, \mathbf{pk})$ takes as input 
		two equal length vectors, a vector of public keys $\mathbf{pk} = (pk_1 , . . . , pk_n)$, a vector of messages $\mathbf{m} = (m_1, . . . , m_n)$, and an aggregate signature $\sigma_\text{agg}$. It outputs either $\true$ or $\false$.
	\end{itemize}
 	\label{def:background:BLS:Correctness}
 	The scheme is correct if for all $\mathbf{pk} = (pk_1, . . . , pk_n), \mathbf{m} = (m_1, . . . , m_n)$, and $\bm{\sigma} = (\sigma_1, . . . , \sigma_n)$, if $\vfyComp(pk_i, m_i , \sigma_i) = \true$ for $i = 1, . . . , n$ then 
 	\begin{equation*}
 	Pr[\vfyAggComp(\aggComp(\mathbf{pk}, \bm{\sigma}), \mathbf{m}, \mathbf{pk}) = \true] = 1
 	\end{equation*}
\end{definition}

Next, we provide the attack game for \ac{EUF-CMA} of aggregate signatures according to Boneh and Shoup~\cite{bonehGraduateCourseApplied}.

\begin{definition}[Attack game for \ac{EUF-CMA} of agg. signatures~\cite{bonehGraduateCourseApplied}, page 626]\label{def:background:attackGameAggregateSignature} 
	For a given aggregate signature scheme
	$\aggsignature$ with message space $\mathcal{M}$, and a given adversary $\mathcal{A}$, the attack game runs as
	follows:
	\begin{itemize}
		\item The challenger runs $(pk, sk) \xleftarrow{R} \genComp()$ and sends $pk$ to $\mathcal{A}$.
		\item $\mathcal{A}$ queries the challenger. For $i = 1, 2, . . . $, the $i$th signing query is a message
		$m^{(i)} \in \mathcal{M}$. The challenger computes $\sigma^{(i)} \xleftarrow{R} \signComp(m^{(i)}, sk)$, and then gives $\sigma^{(i)}$ to $\mathcal{A}$.
		\item Eventually $\mathcal{A}$ outputs a candidate forgery $(\mathbf{pk}, \mathbf{m}, \sigma_\text{agg})$ \\where $\mathbf{pk} = (pk_1 , . . . , pk_n )$ and $\mathbf{m} = (m_1, . . . , m_n) \in \mathcal{M}^n$.
	\end{itemize}
  The adversary wins the game if the following conditions hold:
	\begin{itemize}
		\item $\vfyAggComp(\sigma_\text{agg}, \mathbf{m}, \mathbf{pk}) = \true$,
		\item there is at least one $j$, $1 \leq j \leq n$, such that (1) $pk_j = pk$, and (2) $\mathcal{A}$ did not issue a signing query for $m_j$, meaning that $m_j \notin \{ m^{(1)}, m^{(2)} , . . .\}$.
	\end{itemize}
	We define $\mathcal{A}$’s advantage with respect to $\aggsignature$, denoted \\$\text{ASIGadv}[\mathcal{A}, \aggsignature]$, as the probability that $\mathcal{A}$
	wins the game. 
\end{definition}

Lastly, we provide the EUF-CMA security definition of aggregate signatures according to Boneh and Shoup~\cite{bonehGraduateCourseApplied}.

\begin{definition}[Security of aggregate signatures~\cite{bonehGraduateCourseApplied}, page 626]\label{def:Background:BLS:security}
	We say that an aggregate signature scheme $\aggsignature$ is secure if for all efficient adversaries $\mathcal{A}$, the quantity $\text{ASIGadv}[\mathcal{A}, \aggsignature]$ is negligible.
\end{definition}

\section{BLS Aggregate Signatures}\label{Appendix:BLS}
Here, we provide the details on BLS aggregate signatures.
BLS aggregate signature are constructed from bilinear pairings, also called bilinear maps. 
A bilinear pairing is an efficiently computable function $e : \mathbb{G}_0 \times \mathbb{G}_1 \rightarrow \mathbb{G}_T$ where $\mathbb{G}_0$, $\mathbb{G}_1$, $ \mathbb{G}_T$ are cyclic groups of prime order $q$ and $e$ satisfies bilinearity and non-de\-ge\-ne\-ra\-ti\-vi\-ty~\cite{bonehGraduateCourseApplied}.
The central property of bilinear pairings which is used to construct BLS signatures is: for all $\alpha, \beta \in \mathbb{Z}_q$ we have
\begin{equation}
e(g_0^\alpha, g_1^\beta) = e(g_0, g_1)^{\alpha \beta} = e(g_0^\beta, g_1^\alpha)
\end{equation}
where $g_0 \in \mathbb{G}_0$ and $g_1 \in \mathbb{G}_1$ are generators.

\subsection{BLS Signatures}

We here present the definition of BLS signatures by Boneh and Shoup~\cite{bonehGraduateCourseApplied}:

\begin{definition}[the BLS signature scheme~\cite{bonehGraduateCourseApplied}, page 621]\label{def:background:BLSSignature}
Let $e : \mathbb{G}_0 \times \mathbb{G}_1 \rightarrow \mathbb{G}_T$ be a pairing where $\mathbb{G}_0, \mathbb{G}_1, \mathbb{G}_T$ are cyclic groups of prime order q, and where $g_0 \in \mathbb{G}_0$ and $g_1 \in \mathbb{G}_1$ are generators. 
Let $H$ be a hash function that maps messages in a finite set $\mathcal{M}$ to elements in $\mathbb{G}_0$.

The BLS signature scheme, denoted $\signature_{BLS} = (\genComp, \signComp, \vfyComp)$, has message space $\mathcal{M}$ and works as
follows:
	\begin{itemize}
		\item $\genComp()$ (key generation algorithm): 
		\begin{itemize}
			\item secret key: $sk \xleftarrow{\text{R}} \Z_q$
			\item public key: $pk \leftarrow g_1^{sk} \in \mathbb{G}_1$.
		\end{itemize}
		\item $\signComp(m, sk)$: To sign a message $m \in \mathcal{M}$ using a secret key $sk \in \Z_q$, do
		$\sigma \leftarrow H(m)^{sk} \in \mathbb{G}_0$
		\item $\vfyComp(\sigma, m, pk)$: To verify a signature $\sigma \in \mathbb{G}_0$ on a message $m \in \mathcal{M}$, using the public key $pk \in \mathbb{G}_1$, output $\true$ if $e(H(m), pk) = e(\sigma, g_1)$
	\end{itemize}
\end{definition}

The BLS signature scheme $\signature_{BLS}$, is proven to be secure under the \ac{co-CDH} assumption~\cite{boneh2001short}. 
Boneh et~al.~\cite{bonehAggregateVerifiablyEncrypted2003a} define this property as follows:
\begin{definition}[\ac{co-CDH}~\cite{bonehAggregateVerifiablyEncrypted2003a}] Given $g_a, g^x_a \in \mathbb{G}_a$ and $h \in \mathbb{G}_b$ compute $h^x \in \mathbb{G}_b$.
\end{definition}

\begin{theorem}[\cite{bonehGraduateCourseApplied}, page 622]
	Let  $e : \mathbb{G}_0 \times \mathbb{G}_1 \rightarrow \mathbb{G}_T$ be a pairing and let $H : \mathcal{M} \rightarrow \mathbb{G}_0$ be a hash function. Then the derived BLS signature scheme $\signature_{BLS}$ is a secure signature scheme assuming \ac{co-CDH} holds for e, and H is modeled as a random oracle. 	
\end{theorem}
For a more formal definition and proof, see~\cite{boneh2001short} and Section~15.5.1 of~\cite{bonehGraduateCourseApplied}.

\subsection{BLS Aggregate Signatures}
BLS signatures support signature aggregation. We will derive the aggregation function by first presenting a naive approach, that is not secure, following the presentation in Boneh and Shoup~\cite{bonehGraduateCourseApplied}.

\begin{definition}[Naive BLS aggregate signature~\cite{bonehGraduateCourseApplied}, page 624]\label{def:background:simpleAggregateBLSConstruction}
	The aggregate signature scheme $\aggsignature_{BLS} = (\signature_{BLS}, \aggComp, \vfyAggComp)$:
	\begin{itemize}
		\item  $\aggComp( \mathbf{pk} \in \mathbb{G}_1^n, \bm{\sigma} \in \mathbb{G}_0^n) := \{\sigma_\text{agg} \leftarrow \sigma_1 \cdot  \sigma_2 \cdots \sigma_n \in  \mathbb{G}_0$, output $\sigma_\text{agg} \in \mathbb{G}_0\}$
		\item $\vfyAggComp(\sigma_\text{agg}, \mathbf{m} \in \mathcal{M}^n, \mathbf{pk} \in \mathbb{G}^n_1) = \true$ if 
		\begin{equation}
		e(\sigma_\text{agg}, g_1) = e (H(m_1), pk_1) \cdots e(H(m_n), pk_n)
		\end{equation}
		
	\end{itemize}
\end{definition}

\label{background:BLSAgg:FasterVerificatoin}
The verification of BLS aggregate signatures can be optimized if all signed messages are identical $m_1 = m_2 = ... =m_n = m$. In that case, the public keys can be aggregated to one aggregate public key $pk_{agg} = pk_1\cdots pk_n \in \mathbb{G}_1$, which is used for the verification:

\begin{equation}
e(\sigma_\text{agg}, g_1) \stackrel{?}{=} e(H(m), pk_\text{agg})
\end{equation}

\noindent This reduces the computation of $n+1$ parings to two pairings. The aggregate public key can even be precomputed and reused.

Note that BLS aggregate signatures cannot be modeled directly in Tamarin using Diffie-Hellman group elements and bilinear pairings, as BLS signature aggregation uses the group operation directly which is not available by the built-in theories that only support exponentiation,
due to the undecidability of unification of such equational theories~\cite{kapurEunificationAlgorithmAnalyzing2003}.

\subsection{Rogue Public Key Attack}\label{appendix:Background:RoguePublicKeyAttack}
As described on a high level in Section~\ref{AttackModels:RogueAttack}, the above naive construction is insecure since it is vulnerable to a rogue public key attack. 
Here, we  provide more details on the rogue public key attack.

The adversary creates the rogue public key $pk_\text{rogue}$ for a target public key $pk_\text{target}$ by choosing a random value $\alpha \xleftarrow{R} \Z_q$ and computing:
\begin{equation}
pk_{rogue} \leftarrow g_1^\alpha / pk_{target} \in \mathbb{G}_1
\end{equation}
The corresponding secret key is $sk_\text{rogue} = \alpha - sk_\text{target}$. However, since the adversary has no access to the target secret key $sk_\text{target}$, the adversary also does not know the rogue secret key $sk_\text{rogue}$. The adversary can still create a valid, rogue signature aggregation:
\begin{equation}
\sigma_{\text{agg}_\text{rogue}} := H(m)^\alpha \in \mathbb{G}_0
\end{equation}
This aggregate signature is valid for the message $m$ repeated twice, the target public key $pk_\text{target}$ and the rogue public key $pk_\text{rogue}$: \\
$\vfyAggComp(\sigma_{\text{agg}_\text{rogue}}, (m, m), (pk_\text{target}, pk_\text{rogue}))$ as:
\begin{align*}
e(\sigma_{\text{agg}_\text{rogue}}, g_1)&=e(H(m)^\alpha, g_1) = e(H(m), g_1^\alpha) \\
&= e(H(m), pk_\text{target} \cdot g_1^\alpha/pk_\text{target})\\
&= e(H(m), pk_\text{target}) \cdot e(H(m), g_1^\alpha/pk_\text{target}) \\
& = e(H(m), pk_\text{target})\cdot e(H(m), pk_{rogue})
\end{align*}

The adversary creates a forgery and thus violates the EUF-CMA security definition of aggregate signatures, Definition~\ref{def:Background:BLS:security}. 
Thus, this naive BLS aggregate signature scheme needs to be extended with mitigations to prevent the rogue public key attack.

\subsection{Mitigating the Rogue Public Key Attack}\label{Appendix:BLS:MitigatingRoguePublicKeyAttack}
There are different methods to prevent the rogue public key attack. We present three of them here.

\subsubsection{Prevent Duplicate Messages}
Note that the rogue public key aggregate $\sigma_{\text{agg}_\text{rogue}}$ validates for the message $m$ repeated twice. In fact, this attack is only possible with multiple times the same message. The first mitigation, presented by Boneh et~al.~\cite{bonehAggregateVerifiablyEncrypted2003a}, addresses this, by enforcing distinct messages. The verification algorithm verifies the distinctness of the messages and otherwise rejects. This is only suitable for applications with unique messages, for example certificate chains.   

\subsubsection{Message Augmentation}
The second method can be applied, if distinct messages are not given. In that case, the distinctness is achieved by prepending the signing public key to every message, before signing. 

\begin{definition}[BLS agg. signature with message augmentation~\cite{bonehGraduateCourseApplied}, page 626]
	Our modified aggregation scheme, denoted $\aggsignature^{(1)}_{BLS}$, is the same as $\aggsignature_{BLS}$ in~\ref{def:background:simpleAggregateBLSConstruction} except that the signing algorithm now uses a hash function $H : G_1 \times \mathcal{M} \rightarrow	\mathbb{G}_0$ and is defined as
	\begin{equation}
	\signComp(m, sk) := H(pk, m)^\alpha \text{ where } sk = \alpha \in \Z_q \text{ and } pk \rightarrow g^\alpha_1
	\end{equation}
	In effect, the message being signed is the pair $(pk, m) \in \mathbb{G}_1 \times \mathcal{M}$. The verification and aggregate verification algorithms are equally modified to hash the pairs $(pk, m)$. Specifically, aggregate verification works as
	\begin{multline}
	\vfyAggComp(\sigma_\text{agg},  \mathbf{m} \in \mathcal{M}^n, \mathbf{pk} \in \mathbb{G}^n_1)=\true \\
	\text{ if } e(\sigma_\text{agg}, g_1) = e(H(pk_1
	, m_1), pk_1) \cdots e(H(pk_n, m_n), pk_n)
	\end{multline}
\end{definition}

As described in Appendix~\ref{background:BLSAgg:FasterVerificatoin}, the naive approach for BLS aggregate signatures of Definition~\ref{def:background:simpleAggregateBLSConstruction} can be verified faster, if all messages are the same. This optimization is not possible with this method. 

\subsubsection{Proof of Possession of the Secret Key}
This method preserves the fast verification described in Appendix~\ref{background:BLSAgg:FasterVerificatoin}. Recall that in the rogue public key attack, the adversary is not in possession of the rogue secret key $pk_\text{rogue}$. We use this fact for this mitigation approach: The signers have to prove the possession of their secret keys, referred to as Proof of Possession.

\begin{definition}[BLS aggregate signature with Proof of Possession~\cite{bonehGraduateCourseApplied}, page 627]
	The modified aggregation scheme, denoted $\aggsignature^{(2)}_{BLS}$, is the same as $\aggsignature_{BLS}$ defined in~\ref{def:background:simpleAggregateBLSConstruction} except that the key generation algorithm also generates a proof $\pi$ to show that the signer has possession of the secret key. We attach this proof $\pi$ to the public key, and it is checked during aggregate verification. In particular, the key generation and aggregate verification algorithms use an auxiliary hash function $H' : \mathbb{G}_1 \rightarrow \mathbb{G}_0$, and operate as follows:
	\begin{itemize}
		\item $\genComp() := \left\{ \begin{array}{rcl}
		\alpha \xleftarrow{R} \Z_q, u \leftarrow g^\alpha_1 \in \mathbb{G}_1, \\\pi \leftarrow H'(u)^\alpha \in \mathbb{G}_0 \\
		\text{ output } pk := (u, \pi) \in \mathbb{G}_1 \times \mathbb{G}_0 \\ \text{ and } sk := \alpha \in \Z_q
		\end{array}\right\}.$
		\item  $\vfyAggComp(\sigma_\text{agg}, \mathbf{m} \in \mathcal{M}^n, \mathbf{pk}):$\\ Let $\mathbf{pk} = (pk_1, . . . , pk_n) = ((u_1, \pi_1), . . . ,(u_n, \pi_n))$	be $n$ public keys, and let $\mathbf{m} = (m_1, . . . , m_n)$. Accept if
		\begin{itemize}
			\item valid proofs: $e(\pi_i, g_1) = e(H'(u_i), u_i)$ for all $i = 1, . . . , n$
      \item and valid aggregate: \\$e(\sigma_\text{agg}, g_1) = e(H(m_1), u_1)\cdots e(H(m_n), u_n)$		
		\end{itemize}		
	\end{itemize}
	The new term $\pi = H'(u)^\alpha$ in the public key is used to prove that the public key owner is in possession of the secret key $\alpha$. This $\pi$ is a BLS signature on the public key $u \in \mathbb{G}_1$, but using the hash function $H'$ instead of $H$. The aggregate verification algorithm first checks that all the terms
	$\pi_1, ..., \pi_n \in \mathbb{G}_0$ in the given public keys are valid, and then verifies that the aggregate signature $\sigma_\text{agg}$	is valid exactly as in $\aggsignature_{BLS}$.
\end{definition}

The above presented schemes $\aggsignature^{(1)}_{BLS}$ and $\aggsignature^{(2)}_{BLS}$ are secure in the sense of Definition~\ref{def:Background:BLS:security}, assuming \ac{co-CDH} holds for $e$, and the hash functions $H$ and $H'$ are modeled as random oracles, see~\cite{bonehAggregateVerifiablyEncrypted2003a} and~\cite{bonehGraduateCourseApplied} Section~15.5.3.3 for the proofs.

\subsection{Attacks on BLS Signatures}
In Section~\ref{sec:background:ZeroSplittingAttack}, we described the splitting zero attack by Quan~\cite{quan2021} on a high level. 
Here we provide more details on this attack.
We first describe the attack on single BLS signatures an then the extension to BLS aggregate signatures.

\subsubsection{Identity Element Keys with a Single BLS Signature}
Quan~\cite{quan2021} first presents an attack on single BLS signatures. 
The adversary chooses the secret key to be equal to zero, this results in the corresponding public key and all signatures signed with it being the identity element: 
\begin{align}
sk_\text{adv} &:= 0\\
pk_\text{adv} &:= g_1^{sk_\text{adv}}=g_1^0 = 1\\
\sigma_\text{adv} &:= H(m)^{sk_\text{adv}} = H(m)^0 = 1
\end{align}
As $\vfyComp(\sigma_\text{adv}, m, pk_\text{adv}) = \true$ if $e(H(m), pk_\text{adv}) = e(\sigma_\text{adv}, g_1)$, and $e(H(m), 1) = e(1, g_1)$ for any message $m \in \mathcal{M}$, the signature $\sigma_\text{adv}$ validates for the public key $pk_\text{adv}$ and any message $m \in \mathcal{M}$.

The BLS IETF draft~\cite{bonehDraftirtfcfrgblssignature04BLSSignatures} requires in the verification algorithm to check that the public key is not the identity element. According to Quan~\cite{quan2021}, some libraries implementing the IETF draft omit this check or implement it incorrectly, which makes this attack practical for those implementations.

\subsubsection{Splitting Zero Attack}\label{section:background:ZeroSplitting}
As described in Section~\ref{sec:background:ZeroSplittingAttack}, Quan~\cite{quan2021} 
extends this attack for BLS aggregate signatures and bypasses the identity element check by choosing two non-honest keys, such that $sk_{\text{mal}_1} + sk_{\text{mal}_2} = 0$. And both $sk_{\text{mal}_1}$ and $sk_{\text{mal}_2} $ are non-zero. 
The adversary then signs some message $m$ with each non-honest key $sk_{\text{mal}_1}$ and $sk_{\text{mal}_2}$. This results in the aggregate of those signatures being the identity element:
\begin{align*}
\sigma_{\text{agg}_\text{mal}} &:= \sigma_{\text{mal}_1} \cdot \sigma_{\text{mal}_2} \\
&= H(m)^{sk_{\text{mal}_1}} \cdot H(m)^{sk_{\text{mal}_2}} \\
&= H(m)^{sk_{\text{mal}_1} + sk_{\text{mal}_2}} \\
&= H(m)^{0}\\
&= 1
\end{align*}

Note that the product of the two non-honest public keys is also the identity element: 
\begin{equation}
pk_{\text{mal}_1} \cdot pk_{\text{mal}_2} = 1
\end{equation}
The adversary can aggregate this non-honest signature $\sigma_{\text{agg}_\text{mal}} = 1$ with some third valid signature $\sigma_3$. This results in an aggregate signature equal to this third signature $\sigma_3$: 
\begin{equation}
\sigma_{\text{agg}_{1,2,3}} := \sigma_{\text{mal}_1} \cdot \sigma_{\text{mal}_2} \cdot \sigma_3 = \sigma_3
\end{equation}
This aggregate signature will validate against the message vector $(m', m', m_3)$ where $m'$ can be any message. We show this in the following: As $\sigma_3$ is a valid signature, we have $e(g_1, \sigma_3) = e(pk_3, H(m_3))$. The following derivation shows, that \\$\vfyAggComp(\sigma_{\text{agg}_{1,2,3}}, (m', m', m_3), (pk_{\text{mal}_1}, pk_{\text{mal}_2}, pk_3))=\true$ for any message $m'$:
\begin{align*}
&e(g_1, \sigma_{\text{agg}_{1,2,3}}) \\=~&e(g_1, \sigma_3)\\
=~&1 \cdot e(pk_3, H(m_3))\\
=~&e(g_1, H(m'))^{0} \cdot e(pk_3, H(m_3))\\
=~&e(g_1, H(m'))^{sk_{\text{mal}_1} + sk_{\text{mal}_2}} \cdot e(pk_3, H(m_3))\\
=~&e(g_1, H(m'))^{sk_{\text{mal}_1}} \cdot e(g_1, H(m'))^{sk_{\text{mal}_2}} \cdot e(pk_3, H(m_3))\\
=~&e(g_1^{sk_{\text{mal}_1}}, H(m')) \cdot e(g_1^{sk_{\text{mal}_2}}, H(m')) \cdot e(pk_3, H(m_3))\\
=~&e(pk_{\text{mal}_1}, H(m')) \cdot e(pk_{\text{mal}_2}, H(m')) \cdot e(pk_3, H(m_3))
\end{align*}

As mentioned in Section~\ref{sec:background:ZeroSplittingAttack}, this attack 
does not violate the EUF-CMA security definition of aggregate signatures, Definition~\ref{def:Background:BLS:security}.
The attack violates uniqueness, which is not guaranteed by EUF-CMA security.
This might be surprising, as single BLS signatures are unique and thus popular in blockchain schemes.
Thus, it is important to note that BLS aggregate signatures are not unique and thus do not provide consensus.
Other primitives or protocols have to provide consensus, this cannot be achieved by aggregate signatures which only provide \ac{EUF-CMA} security.

Interestingly, the first two methods to prevent the rogue public key attack (preventing duplicate messages and message augmentation) will also prevent this attack, since aggregating signatures on the same message will be prevented. 
However, the Proof of Possession method does not resolve this problem, as the adversary is in possession of the two non-honest secret keys.

\label{explanationSplittingZeroMitigationsAreExpensive}As each subset of public keys with the same signed message could be colluded, the attack could only be prevented by checking not only all pairs, but all subsets of public keys. 
For a large number of signatures with the same signed message, this would be quite expensive.

\section{Validation Model Restrictions in Tamarin}\label{Appendix:TamarinRestrictions}
In Section~\ref{Sec:VerifModels:Restrictions}, we provided the restrictions that define the validation models and in Section~\ref{Sec:ValidationModel:AggInTamarin}, we mentioned some aspects on how we formalized these restrictions in Tamarin.
Here we provide more details on this formalization in Tamarin. 
We discuss the correctness restriction as an example.
We reformulate Restriction~\ref{restr:restrictionBased:Correctness} on Page~\pageref{restr:restrictionBased:Correctness} to the following equivalent restriction:
\begin{tamarinrestriction}
	\begin{multline*}
	\forall \mathbf{pk}, \mathbf{m}, \bm{\sigma}. \vfyAgg(\agg(\bm{\sigma}), \mathbf{m}, \mathbf{pk}, \false)\\
	\Rightarrow
	(\exists i \in \{1, ..., n\}. (\lnot\sigma_i = \sign(m_i, sk_i) \lor \lnot\Honest(pk_i)))
	\end{multline*}
\end{tamarinrestriction}

We translate this into the following Tamarin restriction, which we will discuss in detail:

\begin{lstlisting}[language=Tamarin, escapechar=*]
restriction Verification_Correctness_morePrecise: 
"All aggregation mAndPk #i. *\label{line:correctnessRestriction:LeftHandside:First}*
VfyAgg(aggregation, mAndPk, false)@i *\label{line:correctnessRestriction:LeftHandside:Last}*
==> 
((Ex si ind thetaAgg mi ski thetaMPk. *\label{line:correctnessRestriction:RightHandside:Multiple:First}*
	VfyAgg(agg(<si, ind>+thetaAgg)*\label{line:correctnessRestriction:Compare:Multiple}*
		, <mi, pk(ski), ind> + thetaMPk, false)@i 
	& (not(si = sign(mi, ski)) *\label{line:correctnessRestriction:RightHandside:Multiple:signature}*
	| not(Ex #j. RegisterHonestKey(pk(ski))@j)))*\label{line:correctnessRestriction:RightHandside:Multiple:Last}*
|(Ex si ind mi ski. *\label{line:correctnessRestriction:RightHandside:One:First}*
   VfyAgg(agg(<si,ind>),<mi,pk(ski),ind>,false)@i*\label{line:correctnessRestriction:Compare:One}*
	& (not (si = sign(mi, ski)) *\label{line:correctnessRestriction:RightHandside:One:signature}*
	| not(Ex #j. RegisterHonestKey(pk(ski))@j))))" *\label{line:correctnessRestriction:RightHandside:One:Last}*
\end{lstlisting}
Lines~\ref{line:correctnessRestriction:LeftHandside:First} to~\ref{line:correctnessRestriction:LeftHandside:Last} are the left-hand side of the implication.
We express 
$\vfyAgg(\sigma_\text{agg}, \mathbf{m}, \mathbf{pk}, \false)$
with the action fact \texttt{VfyAgg(aggregation, mAndPk, false)}. We quantify over the occurrences of verification with the result $\false$ and over all signature aggregations, messages, and public keys. In other words, for each trace that contains an action fact \texttt{VfyAgg} with the verification result $\false$, the right-hand side expressed in lines~\ref{line:correctnessRestriction:RightHandside:Multiple:First} to~\ref{line:correctnessRestriction:RightHandside:One:Last} must hold.

For the right-hand side, we have to do a case distinction. Lines~\ref{line:correctnessRestriction:RightHandside:Multiple:First} to~\ref{line:correctnessRestriction:RightHandside:Multiple:Last} handle the case of two or more aggregated signatures and lines~\ref{line:correctnessRestriction:RightHandside:One:First} to~\ref{line:correctnessRestriction:RightHandside:One:Last} handle the case of one aggregated signature. 
To explain why we need this distinction, we first explain how we express the existential quantification of the right-hand side. The existential quantification $\exists i \in \{1, ..., n\}$ quantifies over the signatures, messages, and secret keys with the same index. We now look at how we express $\exists \sigma_i \in \bm{\sigma}$. The aggregation $\agg(\bm{\sigma})$ is represented as 
\begin{lstlisting}[language=Tamarin, numbers=none]
agg(<s1, ind1> + ... + <sn, indn>)
\end{lstlisting}
Due to the associative-commutative property of multisets,  this matches 
\begin{lstlisting}[language=Tamarin, numbers=none]
agg(<si, ind> + thetaAgg)
\end{lstlisting}
See the following equation:
\begin{multline*}
\langle\sigma_1, i_1\rangle + \langle\sigma_2, i_2\rangle + .... + \langle\sigma_n, i_n\rangle
\equiv_{AC} \langle\sigma_i, i_i\rangle + \\ \underbrace{\langle\sigma_1, i_1\rangle + ... + \langle\sigma_{i-1}, i_{i-1}\rangle + \langle\sigma_{i+1}, i_{i+1}\rangle + ... + \langle\sigma_n, i_{n}\rangle}_{\vartheta_{agg}}
\end{multline*}
Thus we can express the existential quantification $\exists \sigma_i \in \bm{\sigma}$ with 
\begin{lstlisting}[language=Tamarin, numbers=none]
Ex si ind thetaAgg. 
    aggregation = agg(<si, ind> + thetaAgg)
\end{lstlisting}
where \texttt{thetaAgg} is a variable that can be instantiated by the remaining signatures. Instead of using this equality, we restate the action fact \texttt{VfyAgg} in lines~\ref{line:correctnessRestriction:RightHandside:Multiple:First} and~\ref{line:correctnessRestriction:Compare:Multiple} with \texttt{aggregation} replaced with \texttt{agg(<si, ind>+thetaAgg)}. We discuss these different formulations at the end of this section.

Now back to the case distinction: In our models, we do not use an empty element in the multisets. \texttt{thetaAgg} will be instantiated by a multiset or a tuple \texttt{<si, ind>}. Thus, with \texttt{agg(<si, ind>+thetaAgg)}, we can only represent signature aggregations with two or more elements. Therefore, we need to cover the case of one aggregated signature separately. See line~\ref{line:correctnessRestriction:Compare:One}, where the aggregation of one signature is represented by \texttt{agg(<si, ind>)}. In other words, we need the case distinction, as \texttt{agg(<si, ind>+thetaAgg)} and \texttt{agg(<si, ind>)} do not pattern match. For other restrictions, we added a second restriction to cover the case of one aggregated signature.

The rest of the implication $\lnot\sigma_i = \sign(m_i, sk_i) \lor \lnot\Honest(pk_i)$ can be translated straightforwardly in lines~\ref{line:correctnessRestriction:RightHandside:Multiple:signature}, \ref{line:correctnessRestriction:RightHandside:Multiple:Last}, \ref{line:correctnessRestriction:RightHandside:One:signature}, and~\ref{line:correctnessRestriction:RightHandside:One:Last}.

We noted above, that we can express $\sigma_i \in \bm{\sigma}$ by stating 
\begin{lstlisting}[language=Tamarin, numbers=none]
aggregation = agg(<si, ind>+thetaAgg)
\end{lstlisting} 
or by restating the action fact where \texttt{aggregation} is replaced by \texttt{agg(<si, ind>+thetaAgg)}. Let us look, as an example, at the following two restrictions:

\begin{lstlisting}[language=Tamarin]
restriction OneMessageKeyPairPerSignature_RestateActionFact:
  "All si thetaAgg ind messagesKeys #i. 
    VfyAgg(agg(<si, ind>+thetaAgg), messagesKeys, true)@i
  ==> Ex mi ski thetaMPk. 
    VfyAgg(agg(<si, ind>+thetaAgg)
      , <mi, pk(ski), ind> + thetaMPk, true)@i" 

restriction OneMessageKeyPairPerSignature_Equation:
  "All si thetaAgg ind messagesKeys #i. 
    VfyAgg(agg(<si, ind>+thetaAgg), messagesKeys, true)@i
  ==> Ex mi ski thetaMPk. 
    messagesKeys = <mi, pk(ski), ind> + thetaMPk" 
\end{lstlisting}

The two restrictions are equivalent. They ensure that there is a signature for each message and key pair. Note that the first one restates the action fact \texttt{VfyAgg} after the implication, while the second explicitly equates \texttt{<mi, pk(ski), ind>+thetaMPk} and \texttt{messagesKeys}. 
The two formulations are equivalent. But interestingly, using the first formulation results in non-termination. Exploring how different formulations of the same restriction are treated by Tamarin would be interesting for future work. \label{par:Restriction:TwoFormulationsOfRestriction}

\section{OAS Definitions}
\label{OAS_corr_unforg}

In Section~\ref{sec:SANA:OAS} we provided an overview of the \emph{Optimistic Aggregate Signature} (OAS) scheme by~\cite{ambrosinSANASecureScalable2016}.
Here we provide their definitions.

\begin{definition}[OAS, Adapted from~\cite{ambrosinSANASecureScalable2016}]
	\label{def:OAS}
	An Optimistic Aggregate Signature (OAS) scheme is a quintuple of probabilistic polynomial time algorithms 
	(\oaskgen{}, \oasaggpk{}, \oassgn{}, \oasaggsgn{}, \oasvfy{}).
		Let $l\in\mathbb N$ be the security parameter.
	\begin{itemize}
	  \item \oaskgen$(1^l)\rightarrow (pk, sk)$ generates a secret signing key $sk$ and a public verification key $pk$.
	  \item \oasaggpk$(pk_1, \dots, pk_n) \rightarrow apk$ aggregates public keys $pk_1, \dots, pk_n$ into an aggregate public key $apk$.
	  \item \oassgn$(m, M, sk) \rightarrow \sigma$ generates a signature $\sigma$ given a message $m \in \{0, 1\}^*$, a default message $M\in\{0, 1\}^*$ and a secret key $sk$. 
	  \item \oasaggsgn$(\sigma_1, \sigma_2, M) \rightarrow \sigma_\text{agg}$ aggregates the aggregate signatures $\sigma_1$ and $\sigma_2$ into the aggregate signature $\sigma_\text{agg}$.
	  \item \oasvfy{} is the verification algorithm. Given a set of non-contributing signers $S_\bot$, an aggregate signature $\sigma_\text{agg}$, a default message $M$ and an aggregate public key $apk$ it outputs either $\bot$ if the signature is invalid or a set $\mathcal B = \{(m_i, S_i) | i \in \{1, \dots, \mu\}\}$, 
	  where $S_i$ is the set of public keys $pk_i$ whose corresponding signing key $sk_i$ was used to sign the message $m_i$.   
	\end{itemize}
	  \end{definition}
\begin{definition}[Correctness of an OAS, adapted from~\cite{ambrosinSANASecureScalable2016}] \label{def:OAS:correctness}
An OAS scheme is correct if 
  \begin{itemize}
    \item[(i)] for all $l \in \mathbb N$, all public $(pk, sk) \leftarrow \oaskgen(1^l)$, all $S_\bot$ with $pk \not\in S_\bot$ and all $m, M \in \{0, 1\}^*$ we have \\
    $\Pr[\oasvfy(S_\bot, \oassgn(m, M, sk), M, \\ \oasaggpk(S_\bot \cup \{pk\})) = \mathcal R] = 1$, where $\mathcal R = \emptyset$ if $m = M$ and $\mathcal R = \{m, \{pk\}\}$ if $m \neq M$.
    \item[(ii)] aggregation works. Meaning that for all valid aggregate signatures $\sigma_1, \sigma_2$, all distinct sets $S_1, S_2$, all subsets $S_{1,\bot} \subseteq S_1$, $S_{2,\bot}\subseteq S_2$, $M\in\{0, 1\}^*$, \\
    if $\oasvfy(S_{1, \bot}, \sigma_1, M, \oasaggpk(S_1)) \rightarrow \mathcal B_1$ and \\
    $\oasvfy(S_{2, \bot}, \sigma_2, M, \oasaggpk(S_2)) \rightarrow \mathcal B_2$, then \\
    $\Pr[\oasvfy(S_{1,\bot} \cup S_{2,\bot}, \oasaggsgn(\sigma_1, \sigma_2), M,\\ \oasaggpk(S_1 \cup S_2)) = \mathcal B_1 \sqcup \mathcal B_2] = 1$, 
    where $\mathcal B_1 \sqcup \mathcal B_2$ denotes the set union except that if there exist $m^*, S, S'$ s.t. $(m^*, S)\in\mathcal B_1$ and $(m^*, S') \in \mathcal B_2$, then $\mathcal B_1 \sqcup \mathcal B_2$ contains the element $(m^*, S \cup S')$ instead of the two original tuples $(m^*, S)$ and $(m^*, S')$.
  \end{itemize}
\end{definition}

Informally, OAS unforgeability mandates that no adversary can generate an aggregate signature 
that attributes a message to an honest signer which never signed said message, even if all other signers are compromised. 
\begin{definition}[Unforgeability of OAS, adapted from~\cite{ambrosinSANASecureScalable2016}]
Let $l \in \mathbb{N}$. 
Let \oas{} be an OAS scheme, $\mathcal A$ an adversary and $Q$ the set of messages queried by $\mathcal A$ to its signing oracle $\oassgn(\cdot, sk)$.
We define the unforgeability game for \oas{} as follows: \\

\noindent\small
$(pk, sk) \gets \oaskgen(1^l)$\\
$\left(\sigma, S_\bot, \left(pk_1, \dots, pk_n\right), \left(sk_1, \dots, sk_n\right)\right) \gets \mathcal{A}^{\oassgn(\cdot, sk)}(pk)$\\
$\textbf{If } \exists i. pk_i \neq pk \land (pk_i, sk_i) \not\in \oaskgen(1^l) \textbf{ then return } 0$\\
$S \gets \{pk_1, \dots, pk_n\}$ \\
$apk \gets \oasaggpk(S)$ \\
$\mathcal B \gets \oasvfy(S_\bot, \sigma, M, apk)$\\
$\textbf{If } S_\bot \not\subseteq S \lor \exists (m_i, S_i) \in \mathcal B. S_i \not\subseteq S \textbf{ then return } 0$ \\
$\textbf{If } \exists (m_i, S_i) \in \mathcal B. pk \in S_i \land m_i \not\in Q \textbf{ then return } 1$ \\
$S_M \gets S \setminus \left(S_\bot \cup \bigcup_{(m_i, S_i)\in \mathcal B}S_i\right)$ \\
$\textbf{If }pk \in S_M \land M \not\in Q \textbf{ then return } 1 \textbf{ else return } 0$ \\
\normalsize

\noindent \oas{} is unforgeable if for all polynomial-time adversaries $\mathcal{A}$, the probability that the unforgeability game outputs $1$ is negligible.
\end{definition}
\section{Attacks on SANA}
\label{SANA:attestation:attacks}\label{Appendix:SANA:Attacks}

In this section, we provide the message sequence charts and some details on the attacks targeting SANA, described in Section~\ref{sec:SANA:Results}. We first present the attacks targeting only the \tokenrequest{} subprotocol, and second we present attacks targeting the complete SANA protocol.

\subsection{Attacks on Token Request}

We provide the three counterexamples corresponding to the weakest notions of authenticity 
the protocol fails to satisfy, namely, Aliveness from the verifier's perspective where the verifier is initialized only with its secret key, and  Weak Agreement from the perspective of the verifier and owner which applies to both initialization scenarios.

\begin{figure}
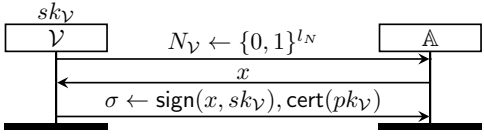

	\centering
	\scalebox{0.9}{
	\begin{msc}[label distance=0.04cm, msc keyword=, instance distance = 4cm, environment distance = 1.0cm, right environment distance=0cm, left environment distance=0cm, last level height=0.1cm, foot height=0.1cm, foot distance=0cm, head height=0.4cm]{}
	  \setlength{\instwidth}{1.5cm}
	  \setlength{\topheaddist}{0.2cm}
	  \setlength{\firstlevelheight}{0.1cm}
	  \drawframe{no}
	  \declinst[label distance=0.1cm]{V}{$\skV$}{$\V$}
	  \declinst{A}{}{$\mathbb A$}
	  \mess{$N_\V \gets \{0, 1\}^{l_N}$}{V}{A}
	  \nextlevel[0.7]
	  \mess{$x$}{A}{V}
	  \nextlevel[1]
	  \mess{$\sigma \gets \signComp(x, \skV), \certComp(\pkV)$}{V}{A}
	\end{msc}
	}
	\caption{The signing oracle $\textsf{SO}(\V, x)$ provides a signature over an arbitrary term $x$ under the secret key of any verifier $\V$.}  
	\label{fig:signing_oracle}
\end{figure}

The attack on Aliveness from the perspective of the verifier relies on the \tokenrequest's signing oracle, provided in Figure~\ref{fig:signing_oracle}.
We introduce the notation $\textsf{SO}(\V, x)$ 
to represent an invocation of this oracle to obtain a signature 
over the term $x$ under $\V$'s secret key $\skV$. 

We provide the counterexample for Aliveness from the verifier's perspective where the verifier is initialized with only it's secret key in Figure \ref{fig:counterexample_aliveness}.
The attack is described in Section~\ref{sec:SANA:Results}.
This attack forms the basis of our attacks on the complete protocol, assuming the verifier is \emph{not} initialized with the owner identity.

\begin{figure}
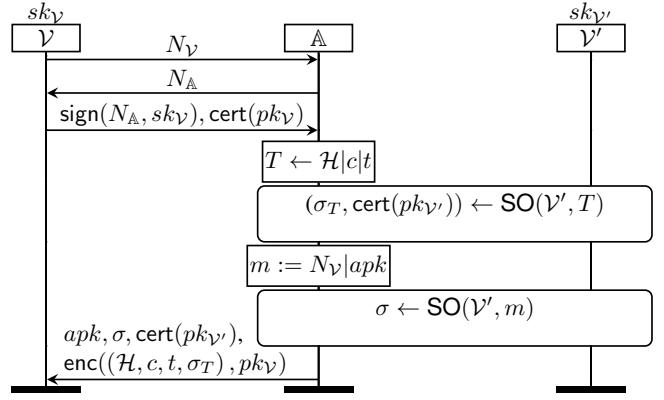

	\centering
	\scalebox{0.9}{
	\begin{msc}[label distance=0.04cm, msc keyword=, instance distance = 3cm, environment distance = 1.5cm, right environment distance=0cm, left environment distance=0cm, last level height=0.1cm, foot height=0.1cm, foot distance=0cm, level height=0.55cm, head height=0.4cm]{}
	  \setlength{\instwidth}{1cm}
	  \setlength{\topheaddist}{0.2cm}
	  \setlength{\firstlevelheight}{0.05cm}
	  \drawframe{no}
	  \declinst[label distance=0.1cm]{V}{$\skV$}{$\V$}
		  \declinst{A}{}{$\mathbb A$}
		  \declinst[label distance=0.1cm]{V2}{$sk_{\V'}$}{$\V'$}
		  		  		  	  \nextlevel[0.1]
		  \mess{$N_\V$}{V}{A}
		  \nextlevel[0.9]
		  \mess{$N_\mathbb{A}$}{A}{V}
		  \nextlevel[1]
		  \mess{$\signComp(N_{\mathbb A}, \skV), \certComp(\pkV)$}{V}{A}
		  \nextlevel[0.3]
	  \action*{$T \gets \mathcal{H} | c | t$}{A}
		  \nextlevel[1.2]
	  \referencestart[left reference overlap=0.9cm, right reference overlap=0.9cm]{T}{($\sigma_T, \certComp(pk_{\V'})) \gets \textsf{SO}(\V', T)$}{A}{V2}
		  \nextlevel[1.5]
		  \referenceend{T}
		  \nextlevel[0.1]
		  \action*{$m := N_\V | apk$}{A}
		  \nextlevel[1.2]
		  		  		  	  \referencestart[left reference overlap=0.9cm, right reference overlap=0.9cm]{s1}{$\sigma \gets \textsf{SO}(\V', m)$}{A}{V2}
		  \nextlevel[1.5]
		  \referenceend{s1}
		  \nextlevel[0.9]
		  \mess{\parbox{3.5 cm}{$apk, \sigma, \certComp(pk_{\V'})$,\\
	  $\encComp(\left(\mathcal{H}, c, t, \sigma_T\right), \pkV)$}}{A}{V}
	\end{msc}
	}
	\caption{Counterexample for the Aliveness property from the perspective of the verifier $\V$. 
  		} 
	\label{fig:counterexample_aliveness}
  \end{figure}

We provide the attack on the Weak Agreement property from the perspective of the verifier in Figure~\ref{fig:counterexample_weakagreement_verifier}.
After a verifier initiates a protocol run with the owner,
all further messages are intercepted by the adversary,
which first finishes the execution with the owner and, thus,
obtains a token encrypted under the adversary's public key
and a signature generated under the owner's secret key.
The adversary can then re-encrypt said token under the verifier's public key
and send it back to the verifier together with the owner's signature.
Therefore, the verifier must believe that it
communicated with the owner, while the latter
actually issued a token to the adversary.

\begin{figure}
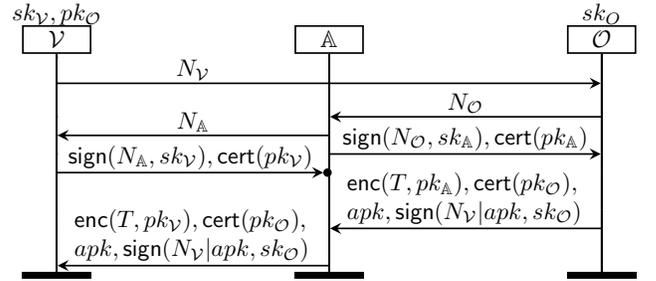

	\centering
	\scalebox{0.9}{
	\begin{msc}[label distance=0.04cm, msc keyword=, instance distance = 3cm, environment distance = 1.5cm, right environment distance=0cm, left environment distance=0cm, last level height=0.1cm, foot height=0.1cm, foot distance=0cm, level height=0.55cm, head height=0.4cm, lost symbol radius = 0.05cm]{}
	  \selfmesswidth = 4cm
	  \setlength{\instwidth}{1cm}
	  \setlength{\topheaddist}{0.2cm}
	  \setlength{\firstlevelheight}{0.05cm}
	  \drawframe{no}
	  \declinst[label distance=0.1cm]{V}{$\skV, \pkO$}{$\V$}
		  \declinst{A}{}{$\mathbb A$}
		  \declinst[label distance=0.1cm]{O}{$sk_O$}{$\Ow$}
		  		  		  	  \nextlevel[0.7]
		  \mess{$N_\V$}{V}[0.25]{O}
	  \nextlevel[0.9]
		  \mess{$N_\Ow$}{O}{A}
		  \nextlevel[0.5]
		  \mess{$N_\mathbb{A}$}{A}{V}
		  \nextlevel[0.5]
		  \mess{$\signComp(N_\Ow, sk_{\mathbb A}), \certComp(pk_{\mathbb A})$}{A}{O}
	  \nextlevel[0.5]
		  \lost[side=right]{$\signComp(N_{\mathbb A}, \skV), \certComp(\pkV)$}{}{V}
	  \nextlevel[1.5]
	  \mess{\parbox{3.5cm}{$\encComp(T, pk_{\mathbb{A}}), \certComp(\pkO),$ \\ $ apk, \signComp(N_\V | apk, \skO)$}}{O}{A}
	  \nextlevel[1]
	  \mess{\parbox{3.5cm}{$\encComp(T, \pkV), \certComp(\pkO),$ \\ $ apk, \signComp(N_\V | apk, \skO)$}}{A}{V}
	\end{msc}
	}
	\caption{Counterexample for the Weak Agreement property from the perspective of the verifier $\V$. 
  		} 
	\label{fig:counterexample_weakagreement_verifier}
\end{figure}

We provide the counterexample for Weak Agreement from the network owner's perspective in Figure~\ref{fig:counterexample_weak_agreement}.
In this attack, the verifier $\V$ starts the \tokenrequest{} with owner $\Ow_2$ while the adversary forwards the verifier's messages to owner $\Ow_1$. 
Thus, the verifier $\V$ and the owner $\Ow_1$ do not agree on their communication partner which contradicts Weak Agreement.

\begin{figure}
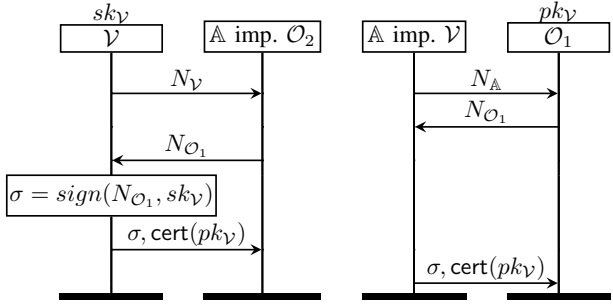

	\centering
	\scalebox{0.9}{
		\begin{msc}[label distance=0.04cm, msc keyword=, environment distance = 1.5cm, right environment distance=0cm, left environment distance=0cm, last level height=0.15cm, foot height=0.1cm, foot distance=0cm, level height=0.55cm, head height=0.4cm]{}
		\setlength{\instwidth}{1.5cm}
		\setlength{\topheaddist}{0.2cm}
				\drawframe{no}
		\declinst[label distance=0.1cm]{V}{$\skV$}{$\V$}
		\declinst[label distance=0.1cm]{O2}{}{$\mathbb{A}$ imp. $\Ow_2$}
		\declinst{A}{}{$\mathbb{A}$ imp. $\V$}
		\declinst[label distance=0.1cm]{O}{$\pkV$}{$\Ow_1$}
		\mess{$N_\V$}{V}{O2}
		\mess{$N_{\mathbb A}$}{A}{O}
		\nextlevel[0.9]
		\mess{$N_{\Ow_1}$}{O}{A}
		\nextlevel[0.9]
		\mess{$N_{\Ow_1}$}{O2}{V}
								\nextlevel[0.4]
		\action*{$\sigma = sign(N_{\Ow_1}, \skV)$}{V}
		\nextlevel[2]
		\mess{$\sigma, \certComp(\pkV)$}{V}{O2}
		\nextlevel[0.9]
		\mess{$\sigma, \certComp(\pkV)$}{A}{O}
		\end{msc}
	}
	\caption{Counterexample for the Weak Agreement property from the perspective of the network owner $\Ow_1$. $\mathbb{A}$ impersonating $\mathcal{X}$ is shortened to $\mathbb{A}$ imp. $\mathcal{X}$}  
	\label{fig:counterexample_weak_agreement}
\end{figure}

\subsection{Attacks on the Complete Protocol}

In this section, we provide the attacks on the complete protocol targeting Attestation Agreement. 
First, we present the attack based on the signing oracle, assuming the verifier is initialized only with its secret key.
Second, we present the attack assuming the verifier is initialized with the owner's identity but assuming the adversary can add rogue public keys to the aggregate public key.

\begin{figure}
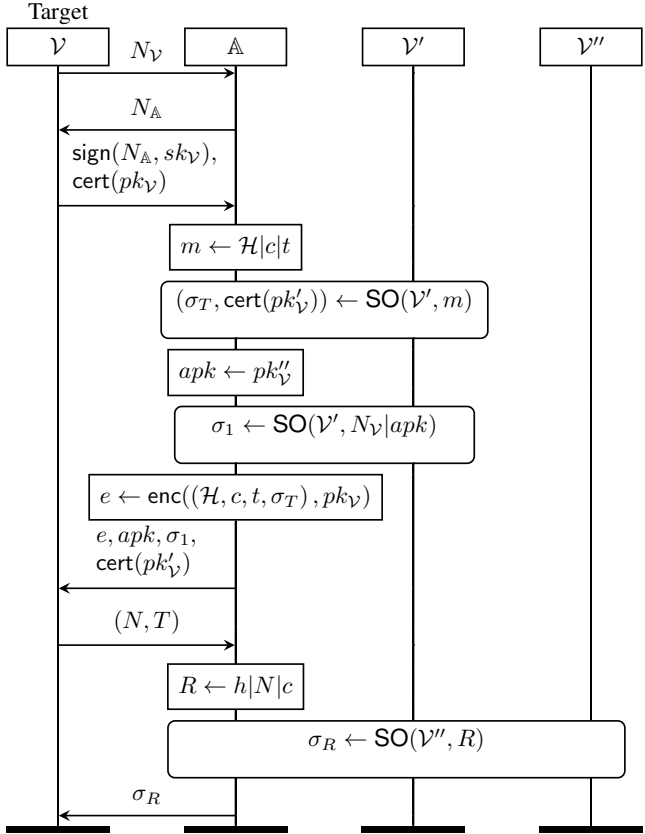

	\scalebox{0.9}{
	  \begin{small}
		\begin{msc}[msc keyword=, instance distance = 1.1cm, environment distance = 1.0cm, right environment distance=0cm, left environment distance=0cm, last level height=0.15cm, foot height=0.1cm, foot distance=0cm, instance width=1cm, level height=0.56cm]{}
		\setlength{\instwidth}{1.5cm}
		\setlength{\topheaddist}{0.2cm}
		\setlength{\firstlevelheight}{0.1cm}
		\drawframe{no}
			\declinst{V}{Target}{$\V$}
			\declinst{A}{}{$\mathbb{A}$}
			\declinst{V1}{}{$\V'$}
			\declinst{V3}{}{$\V''$}
			\mess{$N_\V$}{V}{A}
			\nextlevel[0.5]
		\nextlevel[1]
			\mess{$N_\mathbb{A}$}{A}{V}
			\nextlevel[2]
			\mess{\parbox{2.2 cm}{$\signComp(N_\mathbb{A}, \skV),$\\$\certComp(\pkV)$}}{V}{A}
						\nextlevel[0.5]
			\action*{$m \gets \mathcal H | c | t$}{A}
		\nextlevel[1.5]
		\referencestart[left reference overlap=1.1cm, right reference overlap=1.1cm]{T}{$(\sigma_T, \certComp(\pkV'))\gets \textsf{SO}(\V', m)$}{A}{V1}
			\nextlevel[1.5]
			\referenceend{T}
			\nextlevel[0.3]
			\action*{$apk \gets \pkV''$}{A}
			\nextlevel[1.5]
			\referencestart[left reference overlap=0.9cm, right reference overlap=0.9cm]{a1}{$\sigma_1 \gets \textsf{SO}(\V', N_\V | apk)$}{A}{V1}
			\nextlevel[1.5]
			\referenceend{a1}
			\nextlevel[0.3]
			\action*{$e \gets \encComp(\left(\mathcal H, c, t, \sigma_T \right), \pkV)$}{A}
			\nextlevel[3]
			\mess{\parbox{1.5 cm}{$e, apk, \sigma_1,$\\$\certComp(\pkV')$}}{A}{V}
			\nextlevel[1.5]
			\mess{$(N, T)$}{V}{A}
			\nextlevel[0.5]
			\action*{$R \gets h | N | c$}{A}
			\nextlevel[1.5]
		\referencestart[left reference overlap=1cm, right reference overlap=0.5cm]{ar}{$\sigma_R \gets \textsf{SO}(\V'', R)$}{A}{V3}
			\nextlevel[1.5]
			\referenceend{ar}
			\nextlevel
			\mess{$\sigma_R$}{A}{V}
		\end{msc}
		\end{small}
	}
	\caption{Counterexample for the Attestation Agreement property, assuming no verifier initialization.}
	\label{fig:SANA:attack_uf}
\end{figure}

The first attack is depicted in Figure~\ref{fig:SANA:attack_uf}.
The adversary exploits the signing oracle in the \tokenrequest{} subprotocol (see Figure \ref{fig:signing_oracle}) 
to obtain signatures for both a fake authorization token and one or more fake attestation responses.
Note that neither the network owner nor any of the provers need to be active at any time. 
The attack is solely executed between the adversary, the target verifier, and a few supporting verifiers as signing oracles.

This attack enables the adversary to present an arbitrary attestation result to the targeted verifier, i.e.,
it can make the targeted verifier believe that there are no devices with invalid software in the network while the attestation on the potentially compromised provers was never carried out.
This distorts not only the verifier's perception of the devices' state
but also of what devices even exist.
The aggregate public key should inform the verifier of the list of provers in the network.
However, as the adversary can forge this, it can 
provide the verifier with other verifier's public keys instead of the prover's public keys.

Regarding the practicality of this attack, we highlight two considerations.
First, each invocation of the signing oracle requires the corresponding verifier to initiate a protocol run.
However, some signatures can be obtained offline as the corresponding messages are known beforehand.
Also, 
potential mechanisms causing a verifier to re-run the protocol upon failure may be exploited for repeated oracle access.
Second, the attack relies on the assumption that the signatures from \tokenrequest{} can be aggregated in the OAS scheme.
This is possible if BLS signatures~\cite{bonehShortSignaturesWeil2001} are used,\footnote{The use of the OAS scheme in \tokenrequest{} is explicitly excluded by~\cite{ambrosinSANASecureScalable2016}.} as outlined in Appendix \ref{sec:BLS_as_OAS}.
Since SANA targets IoT devices, which often require small code bases, it is likely that developers choose BLS signatures with an OAS extension instead of two independent signature libraries, making this assumption realistic.

\begin{figure}
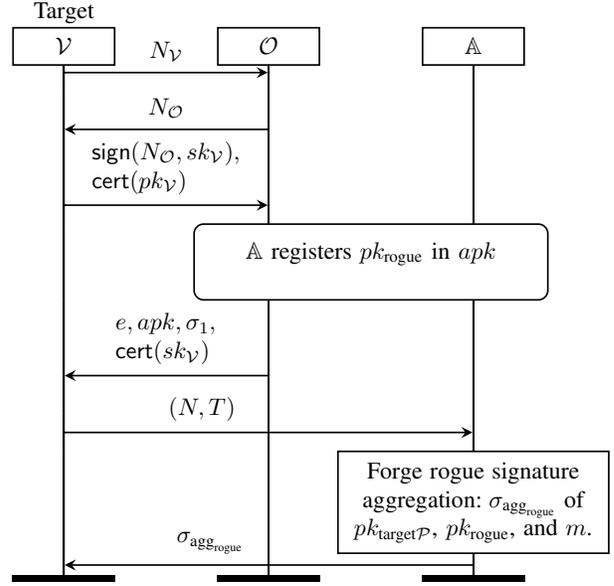

	\scalebox{0.9}{
	  \begin{small}
		\begin{msc}[msc keyword=, instance distance = 1.5cm, environment distance = 1.0cm, right environment distance=0cm, left environment distance=0cm, last level height=0.15cm, foot height=0.1cm, foot distance=0cm, instance width=1cm, level height=0.56cm, action width = 4cm]{}
		\setlength{\instwidth}{1.5cm}
		\setlength{\topheaddist}{0.2cm}
		\setlength{\firstlevelheight}{0.1cm}
		\drawframe{no}
			\declinst{V}{Target}{$\V$}
			\declinst{O}{}{$\Ow$}
			\declinst{A}{}{$\mathbb{A}$}
			\mess{$N_\V$}{V}{O}
			\nextlevel[0.5]
			\nextlevel[1]
			\mess{$N_\Ow$}{O}{V}
			\nextlevel[2]
			\mess{\parbox{2.2 cm}{$\signComp(N_\Ow, \skV),$\\$\certComp(\pkV)$}}{V}{O}
						\nextlevel[0.5]
			\referencestart[left reference overlap=1.1cm, right reference overlap=1.1cm]{T}{$\mathbb{A}$ registers $pk_{\text{rogue}}$ in $apk$}{O}{A}
			\nextlevel[2]
			\referenceend{T}
			\nextlevel[2]
			\mess{\parbox{1.5 cm}{$e, apk, \sigma_1,$\\$\certComp(\skV)$}}{O}{V}
			\nextlevel[1.5]
			\mess{$(N, T)$~~~~~~~~~~~~~~~~~}{V}{A}
			\nextlevel[0.5]
			\action{Forge rogue signature aggregation: $\sigma_{\text{agg}_\text{rogue}}$ of $pk_{\text{target}\mathcal{P}}$, $pk_\text{rogue}$, and $m$.}{A}
			\nextlevel[3]
			\mess{$\sigma_{\text{agg}_\text{rogue}}$~~~~~~~~~~~~~~~}{A}{V}
		\end{msc}
		\end{small}
	}
	\caption{Counterexample for the Attestation Agreement property, assuming rogue public key registration.}
	\label{fig:SANA:attack_rogue}
\end{figure}

The attack, assuming the adversary capability of registering rogue public keys in the aggregate public key, is depicted in Figure~\ref{fig:SANA:attack_rogue}.
The \tokenrequest{} is performed normally, while the attestation response is forged by the adversary.
The adversary registers a rogue public key $pk_\text{rogue}$ for a target prover's public key $pk_{\text{target}\mathcal{P}}$
and provides the rogue signature aggregation $\sigma_{\text{agg}_\text{rogue}}$ of these keys and message $m$ to the verifier.
This forges the target prover's attestation response.

\section{Aggregating BLS Signatures and OAS}
\label{sec:BLS_as_OAS}
In this section, we show how a valid BLS signature can be transformed into a valid OAS signature, which can then be further aggregated into an aggregate OAS signature.
We begin by repeating the concrete OAS scheme given in~\cite{ambrosinSANASecureScalable2016}.

\begin{definition}[Concrete OAS scheme construction according to~\cite{ambrosinSANASecureScalable2016}]
Let $\mathbb G_0, \mathbb G_1$ and $\mathbb G_T$ be multiplicative groups of prime order $p$ with generators $g_0, g_1$ and $g_t$ and let $e : \mathbb G_0 \times \mathbb G_1 \rightarrow \mathbb G_T$ be an efficiently computable bilinear map.
Let $H: \{0,1\}^* \rightarrow \mathbb G_1$ be a hash function.

\begin{LaTeXdescription}
	\item[Key Generation] Secret keys are sampled uniformly random: $sk \gets_\$ \mathbb{Z}_p$.
	Then, $pk \gets g_1^{sk}$.
	\item[Public Key Aggregation] Aggregating public keys $pk_1, \dots, pk_n$ is done via multiplication: $apk = \prod_{i=1}^n pk_i$.
	\item[Signing] Let $m \in \{0, 1\}^*$. Then, $\tau \gets H(m)^{sk}$. If $m$ is the default message, then $\sigma = \{\tau, \emptyset\}$.
	Otherwise, $\sigma = \{\tau, \{(m, \{pk\})\}\}$.    
	\item[Signature Aggregation] Let $\sigma_1 = \{\tau_1, \mathcal B_1\}$ and $\sigma_2 = \{\tau_2, \mathcal B_2\}$.
	Then, $\sigma_1$ and $\sigma_2$ are aggregated as follows: $\sigma \gets \{\tau_1 \cdot \tau_2, \mathcal B_1 \sqcup \mathcal B_2\}$, where $\sqcup$ is defined as in Definition~\ref{def:OAS:correctness}, Appendix~\ref{OAS_corr_unforg}.
	\item[Verification] Let $S_\bot$ be the set of non-contributing public keys, $apk$ the aggregate public key, $M$ the default message and $\sigma = (\tau, \{(m_1, S_1), \dots, (m_\mu, S_\mu)\})$ the aggregate signature.
	Let, 
	$$apk_M \gets apk \cdot \left(\prod_{pk\in S_\bot}pk \cdot \prod_{i = 1}^{\mu}\prod_{pk\in S_i}pk\right)^{-1}$$
	The verification algorithm returns true if
	$$e(\tau, g_1) = e\left(H(M), apk_M\right)\cdot \prod_{i=1}^{\mu}e\left(H(m_i), \prod_{pk\in S_i}pk\right)$$
	and false otherwise.
\end{LaTeXdescription}
\end{definition}

Let \oas{} denote the OAS scheme described above. Let $\mathsf{BLS}$ denote the BLS signature scheme with the same 
groups $\mathbb G_0, \mathbb G_1$ and $\mathbb G_T$, the same bilinear map $e$ and the same hash function $H$.
Let $\sigma_{BLS} = H(m)^{sk}$ be a BLS signature over the message $m$ generated with the key pair $(sk, pk)$ using the algorithms of $\mathsf{BLS}$.
We define $\sigma_{OAS} := (\sigma_{BLS}, \mathcal B)$, where $\mathcal B = \emptyset$ if $m = M$ or $\mathcal B = \{(m, \{pk\})\}$ otherwise.
We now argue that $\oasvfy(\emptyset, \sigma_{OAS}, M, pk) = \true$.
To see this, consider two cases.

If $m = M$, then $apk_M = pk$ and, thus, 
\begin{equation*}
	\begin{split}
		e(\sigma_{BLS}, g_1) \overset{(1)}{=} & e\left(H(m), pk\right) \\
		= & e\left(H(M), apk_M\right).
	\end{split}
\end{equation*}

If $m \neq M$, then $apk_M = 1$ and, therefore,
\begin{equation*}
	\begin{split}
		e(\sigma_{BLS}, g_1) \overset{(1)}{=} & e\left(H(m), pk\right) \cdot 1 \\
		= & e\left(H(m), pk\right) \cdot e\left(H(M), apk_M\right).
	\end{split}
\end{equation*}
In $(1)$ we use the correctness of $\mathsf{BLS}$.
}

\end{document}